\documentclass{aa}
\usepackage{natbib}
\bibpunct{(}{)}{;}{a}{}{,} 
\usepackage{graphicx}
\usepackage{hyperref}
\usepackage{xcolor}
\usepackage{txfonts}
\usepackage{comment}

\usepackage{amsmath}	
\usepackage{orcidlink}
\usepackage{longtable}

\begin{document} 

\title{Distance estimate to NGC~6951 from supernova siblings Type IIP SN~2020dpw and Type Ib SN~2021sjt}

\titlerunning{Supernova siblings SNe~2020dpw 2021sjt in NGC~6951}

\author{
\orcidlink{0000-0002-8770-6764}R\'eka K\"onyves-T\'oth\inst{1,2} \and
Zs\'ofia Bodola\inst{3} \and 
\orcidlink{0000-0003-4610-1117}Tam\'as Szalai\inst{3,4} \and
Moira Andrews \inst{9,10}\and 	
Iair Arcavi \and
\orcidlink{0009-0000-9929-7518} Dominik B{\'a}nhidi\inst{3,5} \and 
Imre Barna B{\'i}r{\'o} \inst{5,6} \and
Federica Bianco\inst{11} \and
Jamison Burke \and 
Istv{\'a}n Cs{\'a}nyi\inst{5} \and
Benjamin Dilday	\and
\orcidlink{0000-0003-4914-5625}Joseph R. Farah\inst{9,10} \and
Estefania Padilla Gonzalez \inst{12}\and
\orcidlink{0000-0002-1125-9187}Daichi Hiramatsu \inst{7}\and
D. Andrew Howell \inst{9,10}\and
Levente Kriskovics\inst{1,2}\and
Megan Newsome \inst{13} \and
\orcidlink{ 0000-0003-3656-5268}Yuan Qi Ni \inst{8,9}\and
Curtis McCully \inst{9} \and
Andr{\'a}s Ordasi\inst{1,2} \and 
Andr{\'a}s P{\'a}l\inst{1,2} \and 
Craig Pellegrino \inst{14} \and
\orcidlink{0000-0002-1698-605X}R{\'o}bert Szak{\'a}ts\inst{1,2}\and 
 Giacomo Terreran\inst{15} \and
Stefano Valenti\inst{16} \and
 \orcidlink{0000-0002-6471-8607}Kriszti{\'a}n Vida\inst{1,2} \and
Kathryn Wynn\inst{9,10}
}

\institute{ 
HUN-REN Research Centre for Astronomy and Earth Sciences, Konkoly Observatory, Konkoly Th. M. {\'u}t 15-17., 1121 Budapest, Hungary 
\and 
CSFK, MTA Centre of Excellence, Konkoly Thege Mikl{\'o}s {\'u}t 15-17, 1121 Budapest, Hungary 
\and
Department of Experimental Physics, Institute of Physics, University of Szeged, D{\'o}m t{\'e}r 9, 6720 Szeged, Hungary 
\and
MTA-ELTE Lend\"ulet "Momentum" Milky Way Research Group, Szent Imre H. st. 112, 9700 Szombathely, Hungary 
\and
 Baja Astronomical Observatory of University of Szeged, Szegedi {\'u}t, Kt. 766, 6500 Baja, Hungary 
 \and
  HUN-REN--SZTE Stellar Astrophysics Research Group, Szegedi {\'u}t, Kt. 766, 6500 Baja, Hungary 
  \and 
  Department of Astronomy, University of Florida, 211 Bryant Space Science Center, Gainesville, FL 32611-2055 USA 
  \and 
  Kavli Institute for Theoretical Physics, University of California, Santa Barbara, 552 University Road, Goleta, CA 93106-4030, USA 
  \and 
  Las Cumbres Observatory Global Telescope Network, 6740 Cortona Drive, Suite 102, Goleta, CA 93117, USA 
  \and 
  Department of Physics, University of California, Santa Barbara, Broida Hall, Mail Code 9530, Santa Barbara, CA 93106-9530, USA 
  \and 
  University of Delaware,  210 S College Ave, Newark, DE 19716, USA 
  \and 
  Space Telescope Science Institute, 3700 San Martin Drive, Baltimore, MD 21218, USA 
  \and 
  Department of Astronomy, University of Texas at Austin, 2515 Speedway, Stop C1400, Austin, Texas 78712-1205, USA 
  \and 
  NASA's Goddard Space Flight Center, 8800 Greenbelt Rd, Greenbelt, MD 20771 
  \and 
  Adler Planetarium, 1300 S. DuSable Lake Shore Drive, Chicago, IL 60605 
  \and 
   UC Davis, One Shields Avenue, Davis, CA 95616 
}

\date{Accepted XXX. Received YYY; in original form ZZZ}

  \abstract
   {Supernova (SN) siblings are powerful tools to calibrate and improve distance measurement methods, and to make the systematic uncertainty to distances to their host galaxies considerably lower compared to other techniques.}
   {In this paper we present distance estimates to NGC6951, a galaxy that hosted the Type IIP SN~2020dpw, the Type Ib SN~2021sjt, and three other SNe.}
   {Photometric observations of the two objects were carried out using two 80cm Ritchey-Chretien telescopes located in Hungary, while spectra were obtained from the LCO and the WiseRep database. After data reduction, distances to the studied SNe were inferred.  For the distance estimates, we applied the expanding photosphere method (EPM), which connects the observed angular radius ($\theta$) of a SN to its physical radius, that is related to the velocity of the photosphere ($v_{\rm ph}$). Although the EPM is mostly applied to derive the distance of Type IIP SNe, in the literature, there are several examples to the usage of this technique for either Type IIn or stripped-envelope SNe as well. Therefore, here we make another attempt to infer the distance of the Type Ib SN~2021sjt applying the EPM together with its Type IIP sibling SN~2020dpw. Here, the $\theta$ values in different epochs for each studied supernova were estimated from photometric observations, while $v_{\rm ph}$ was constrained from the modeling of the available spectra using SYN++.}
   {Our analysis resulted in a distance of $25.76 \pm 0.34 (\rm random) \pm 5.51$ (systematic) Mpc and $24.57 \pm 1.27  (\rm random) \pm 4.64$ (systematic) Mpc for SN~2020dpw and SN~2021sjt, respectively. Systematic errors were estimated with respect to the used dilution factor, the interstellar reddening, and the date of the explosion, which was fixed to a value between the last non-detection and the first detection for each object.}
 {The obtained distance values agree with each other and with the literature, which shows the validity of the methods used. This way, new, and maybe improved distance estimates to NGC 6951 were obtained, and the applicability of the EPM for Type Ib SNe was tested.}
   \keywords{supernovae: general -- supernovae: individual: SN~2021sjt, SN~2020dpw, SN~1999el, SN~2000E, SN~2015G
               }
   \maketitle
   \nolinenumbers

\section{Introduction}

Supernovae (SNe) that explode in the same host galaxy - called SN siblings - offer a unique opportunity to improve extragalactic distance measurements and decrease the systematic uncertainties of such estimates. Since SN siblings explode in the same galaxy, their distances have to be essentially the same, independently from the applied distance measurement method. Up to date, several studies have been published, mostly on the distance measurements of Type Ia siblings \citep[see e.g.][]{2020ApJ...895..118B,2020ApJ...896L..13S,2022MNRAS.509.5340B, 2022ApJ...928..103H, 2022A&A...666A..13G,2023A&A...677A.183B,2023ApJ...956..111W,2024MNRAS.527.8015K,2025ApJ...981...97S}, which helped to refine both the distances of their host galaxies and the value of the Hubble-constant. Type Ia SNe are ideal candidates for sibling studies because of their relatively high absolute brightness and their rate. 

However, there are some studies about Type II SN siblings as well. 
Up to date, for example \citet{2012A&A...540A..93V} presented a combined analysis of a IIb -- IIP supernova pair (SN~2005cs and SN2011dh) in order to constrain the distance of their host galaxy (M51), while two IIP supernovae (SN~2004et and SN~2017eaw) that exploded in the Firework galaxy NGC~6946, were studied for similar reasons by \citet{2019ApJ...876...19S}.
\citet{2022MNRAS.511..241G} carried out distance measurement of a pair of a Type Ia and a Type IIP supernova, resulting in a good agreement in the inferred distances.  
Later, \citet{2023A&A...672A.129C} carried out a more complete analysis of 4 pairs of Type IIP supernova siblings that exploded in 4 different host galaxies in order to check the consistency of EPM and SCM. Apart from testing these techniques, they calibrated the distances of the 4 host galaxies. 

In the casa of Type II SNe, distance estimates are usually done by applying methods such as the Expanding Photosphere Method (EPM)
\citep{1974ApJ...193...27K}, which geometrically connects the angular size with the radius of the photosphere, and the empirical standard candle method (SCM) \citep{2002ApJ...566L..63H}.One of the main differences between EPM and SCM is that the latter provides relative distance estimate. On the contrary, EPM is capable of deriving absolute distances, without external calibration, which is one of the greatest advantages of the method. However, EPM is not free from systematic uncertainties, from which the most important is the assumption of a diluted black body and the date of the explosion. Therefore the obtained distances strongly depend on the used model and correction factors. By calculating the distance of sibling SNe  which occurred in the same host galaxy, the effect of random and systematic errors can be tested.

It has to be mentioned that there are several EPM implementations in the literature. Originally, the method was created and firstly used by \citet{Kirshner_1974, 1994ApJ...432...42S} and \citet{1996ApJ...466..911E}.
Initially, it was assumed that the SN emits blackbody radiation, however, since this simplification does not take into account that the continuum can be diluted by the electron scattering and other processes,
the $\xi$ correction factor was introduced to account for the dilution from the blackbody radiation. This correction factor strongly depends on supernova models, therefore it has a high impact on the derived distances. Earlier, e.g. \citet{2001ApJ...558..615H,1996ApJ...466..911E, Dessart_2005, 2015MNRAS.453.2189D} calculated different dilution factors, which were used by e.g. \citet{2018A&A...611A..25G}, who discussed in details the effect of the choice of different $\xi$ values to the estimated distances.

\citet{Takats_2012} created a slightly different EPM technique, in which quasi-bolometric fluxes are used instead of monochromatic fluxes. In this paper, this implementation is applied to the studied objects. 

Last, but not least, an approach called "tailored EPM" was carried out by \citet{2006A&A...447..691D}, in which complex radiative transfer models were calculated and compared to the observed spectra. This precise, but especially time consuming method was applied to Type II SNe by e.g. \citet{2004ApJ...616L..91B, 2006A&A...447..691D,2008ApJ...675..644D}. Later, improvements were made by \citet{2019A&A...621A..29V,2020A&A...633A..88V} and \citet{2025A&A...702A..41V}
in order to shorten the process: they introduced a spectral emulator, which can interpolate the radiative transfer models in a given parameter space. Apart from \citet{2025A&A...702A..41V}
the tailored EPM was tested  by \citet{2023A&A...672A.129C} as well on Type II sibling supernovae.

In this paper we present new distance calculations to NCG~6951, applying the EPM to two sibling SNe exploded there: the Type IIP SN~2020dpw and the Type Ib SN~2021sjt. It is important to note that SN~2020dpw and SN~2021sjt are not the only supernovae that have exploded in NGC~6951: over the past few decades, three additional supernovae have occurred in the galaxy: SN1999el (IIn), SN2000E (Ia) and SN~2015G (Ibn). In the literature, several attempts were made to calculate the distance to their host galaxy using different methods. \citet{2001AA...372..824V} obtained a distance of 33 $\pm$ 8 Mpc to NGC6951 from the analysis of SN~2000E. They applied the Multi-color Light Curve Shape (MLCS-2) method, which connects the peak brightness of SNe-Ia to the shape of their light curves.  \citet{2003ApJ...595..779V} estimated the distance of SN~2000E as well, and found  D~=~$26.8  \pm 2.6$ Mpc using the Phillips-relation. 
Later, the calculation of \citet{2017MNRAS.471.4381S}, who adopted D~=~$23.1 \pm 3.5$ Mpc to SN~2015G using the average of the distance values published in the NASA/IPAC Extragalactic Database\footnote{ned.ipac.caltech.edu} to the host galaxy.

\begin{figure}
\centering
\includegraphics[width=8cm]{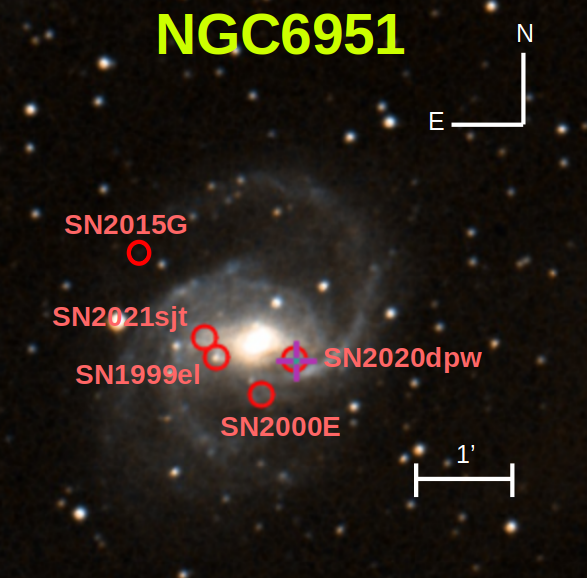}
\caption{The position of supernova siblings SNe~1999el, 2000E, 2015G, 2020dpw, and 2021sjt in NGC~6951   (\url{http://simbad.u-strasbg.fr}). }
\label{fig:siblings}
\end{figure}

In Figure \ref{fig:siblings} an image of NGC~6951, and the position of its 5 "supernova children" can be seen, while Table \ref{tab:irodalom}
collects some basic data of these SNe. The reason for studying only SN~2020dpw and SN~2021sjt is the lack of quality data on SN~1999el, and the inapplicability of the expanding photosphere method on the Type Ia SN~2000E and the Type Ibn SN~2015G. The distance of the former SN can be estimated using methods specialized to Type Ia SNe, while the latter object exploded into the dense circumstellar material, which excludes the application of most direct distance measurement techniques.

\begin{table*}
\caption{Basic data and distance estimates of the SNe that have exploded in NGC~6951.}
\label{tab:irodalom}
\begin{center}
\begin{tabular}{lcccc}
\hline
\hline
Object & R.A. & Dec. & Type  & Distance estimates \\
\hline
SN~1999el & 20:37:17.719 & +66:06:11.48 & IIn & No information \\
\hline
SN~2000E & 20:37:13.769 & +66:05:50.21 & Ia & $33 \pm 8$ Mpc \citet{2001AA...372..824V}\\
         &                &            &     & $26.8  \pm 2.6$ Mpc \citet{2003ApJ...595..779V} \\
\hline
\hline
SN~2020dpw & 20:37:10.550 & +66:06:10.66 & IIP & $23.1 \pm 3.5$ Mpc \citet{2024ApJ...964..172B} from NED\\
            &             &               &     & $25.76 \pm 0.34 \rm(random) \pm 5.51$(systematic) Mpc This paper \\                   
\hline 
SN~2021sjt & 20:37:19.202 & +66:06:23.10 & Ib & $24.57 \pm 1.27\rm(random)\pm4.64$(systematic) Mpc This paper \\
\hline
\end{tabular}
\end{center}
\end{table*}

This paper is structured as follows: Section \ref{sec:obs}
gives information on the observations and used data of SN~2020dpw and SN~2021sjt. Section \ref{sec:epm}
describes the  used version of the expanding photosphere technique. Its application to the two studied SNe is detailed in Section \ref{sec:distances} together with the discussion of the potential sources of systematic error, and their effect to the derived distances, while Section \ref{sec:conc} summarizes the conclusions of the distance measurements of the two SN siblings.

\section{Data}\label{sec:obs}

\subsection{Photometry}
Photometric data of the two studied SNe were obtained from several sources. 

SN~2020dpw was observed using the 0.8m Ritchy--Chrétien telescope of Konkoly Observatory, Hungary (RC80). RC80 was manufactured by the AstroSysteme Austria, and is equipped with a back-illuminated, 2048 $\times$ 2048 pixel FLI PL230 CCD chip having 0.55" pixel scale, and Johnson--Cousins {\it BV} and Sloan {\it griz} filters. With this telescope, photometric observations of nearby ($z < 0.1$) supernovae can be obtained with a signal-to-noise (S/N) ratio of $\gtrsim 10$.
 
{\it BVgri} photometry of the other studied target, SN~2021sjt was performed by the twin of RC80 located at the Baja Observatory of the University of Szeged, Hungary (called BRC80).

RC80 and BRC80 data were processed with standard  Image Reduction and Analysis Facility (IRAF) \citep{1993ASPC...52..173T,1986SPIE..627..733T}
routines, including basic corrections. Then we co-added three images per filter per night aligned with the {\tt wcsxymatch}, {\tt geomap} and {\tt geotran} tasks. We obtained aperture photometry on the co-added frames using the {\tt daophot} package in IRAF, and image subtraction photometry based on other IRAF tasks (e.g.,  {\tt psfmatch} and {\tt linmatch}, respectively). For the image subtraction we applied a template image taken at a sufficiently late phase, when the transient was no longer detectable on our frames.

Moreover, further {\it BVgri} data of SN~2021sjt were collected from Las Cumbres Observatory (LCO) utilizing a world-wide network of telescopes under the Global Supernova Project \citep[GSP,][]{Howell_2019}. After the basic reduction (bias-, dark- and flatfield-correction) of the available LCO data, the frames taken on the same night with the same filter were median-combined, then aperture photometry was performed using the fitsh software \citep{2012MNRAS.421.1825P}. 

The photometric calibration was carried out using
stars from Data Release 1 of Pan-STARRS1 (PS1 DR1).\footnote{\href{https://catalogs.mast.stsci.edu/panstarrs/}{https://catalogs.mast.stsci.edu/panstarrs/}} 
In order to obtain reference magnitudes for our $B$- and $V$-band frames, the PS1 magnitudes were transformed into the Johnson $BVRI$ system based on equations and coefficients found in \cite{Tonry12}. Finally, the instrumental magnitudes were transformed into standard $BVgri$ magnitudes by applying a linear color term (using $g-i$) and wavelength-dependent zero points. Since the reference stars fell within a few arcminutes around the target, no atmospheric extinction correction was necessary. S-corrections were not applied.

\begin{figure*}[h!]
\centering
\includegraphics[width=8cm]{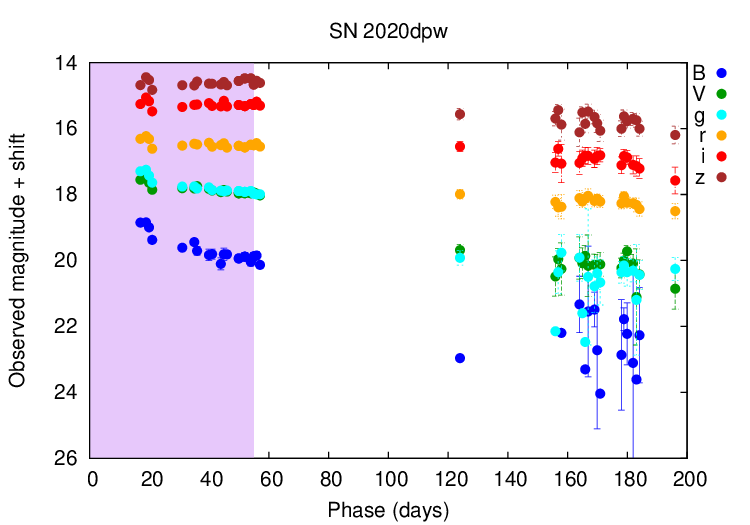}
\includegraphics[width=8cm]{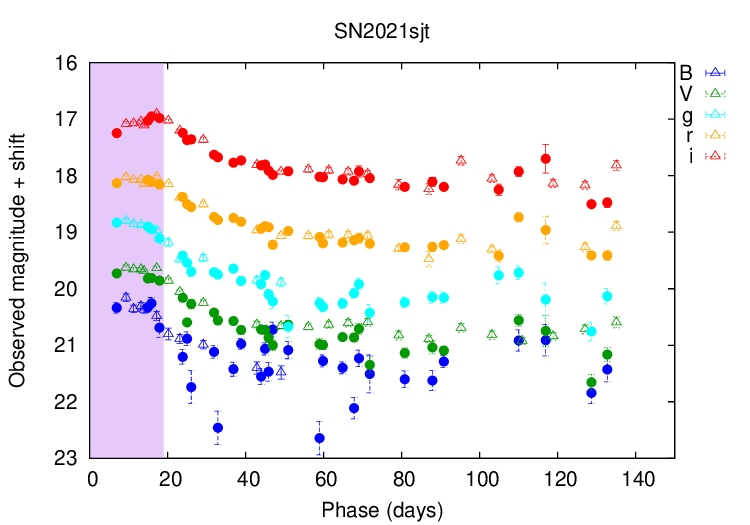}
\caption{Photometric data of SN~2020dpw (left) and SN~2021sjt (right) from different sources (see Section \ref{sec:obs}). Filled dots denote to RC80 and BRC80 data, respectively, while empty triangles code the LCO data. The purple-shaded regions in each panel refer to the time interval, where the expanding photosphere method was applied (see its explanation in \ref{sec:distances}.  }
\label{fig:lcs}
\end{figure*}

The resulting $BVgriz$ and $BVgri$ light curves of SN~2020dpw and SN~2021sjt, respectively, can be seen in Figure \ref{fig:lcs}. 
The photometric data of the two studied SNe are collected in Tables \ref{tab:20dpw_lc},  \ref{tab:21sjt_lc_brc} and \ref{tab:21sjt_lc_lco} in the Appendix.)

\subsection{Spectroscopy}\label{sec:spec}

In the case of SN~2020dpw, spectra were downloaded from the WiseRep database \footnote{https://www.wiserep.org} \citep{2012PASP..124..668Y}. Out of the two available spectra, only the second one (obtained at +20.4 days phase post-explosion) has an acceptable S/N -- thus, we used only this one for further analysis. For SN~2021sjt, we obtained 5 optical spectra at the LCO with the FLOYDS spectrograph mounted on the 2m
Faulkes Telescope North (FTN) at Haleakala (USA), through the GSP, between 2021 July 10--25 (between phases of +4 and +19 days post explosion). A 2\arcsec\  wide slit was placed on the target at the parallactic angle \citep{filippenko82}. We extracted, reduced, and calibrated the  1D spectra following the standard procedures using the FLOYDS pipeline \citep{Valenti_2014}.
 
The log of the spectroscopic observations can be found in Table \ref{tab:spectra} in the Appendix.

To make these spectra appropriate for the spectrum modeling, which was necessary to estimate the photoshperic velocities at each observed phase, they were corrected for both redshift and interstellar extinction. For these corrections, a redshift value of $z~=~0.0048$, and a total interstellar reddening (i.e. the sum of the Milky Way reddening and the extinction due to the host galaxy), $E(B-V)_{\rm tot}~=~0.38$ was applied. According to \citet{2011AAS...21743442S}, the Milky Way reddening for both SNe~2020dpw and 2021sjt are 0.32 mag, while we adopted the extinction due to the host galaxy, 0.06 mag, from \cite{Shivvers_2017}, who analyzed the high-resolution spectra of SN~2015G appeared in the same galaxy (such data are not available for either SN 2020dpw, or 2021sjt).

\begin{figure*}[h!]
\centering
\includegraphics[width=6cm]{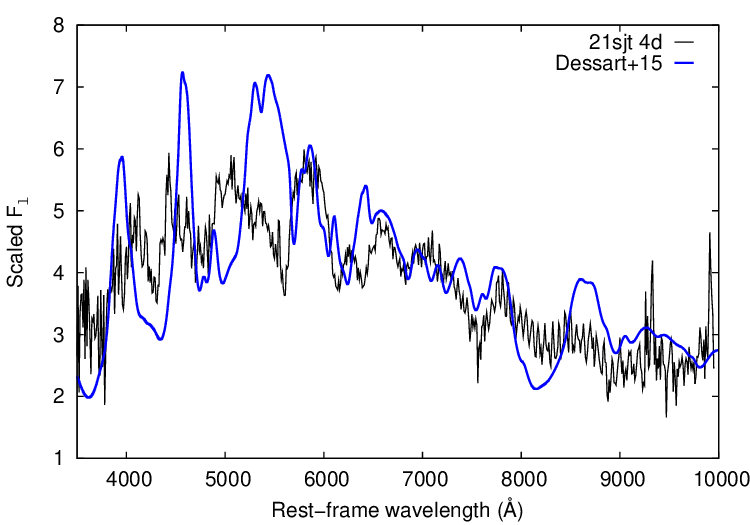}
\includegraphics[width=6cm]{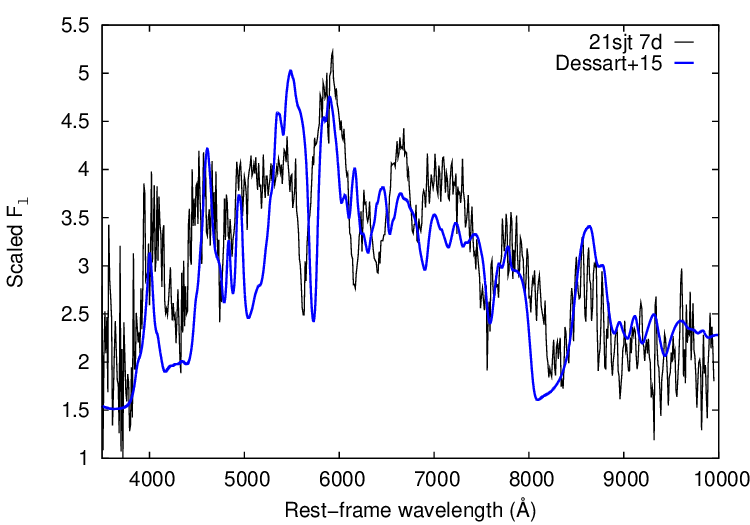}
\includegraphics[width=6cm]{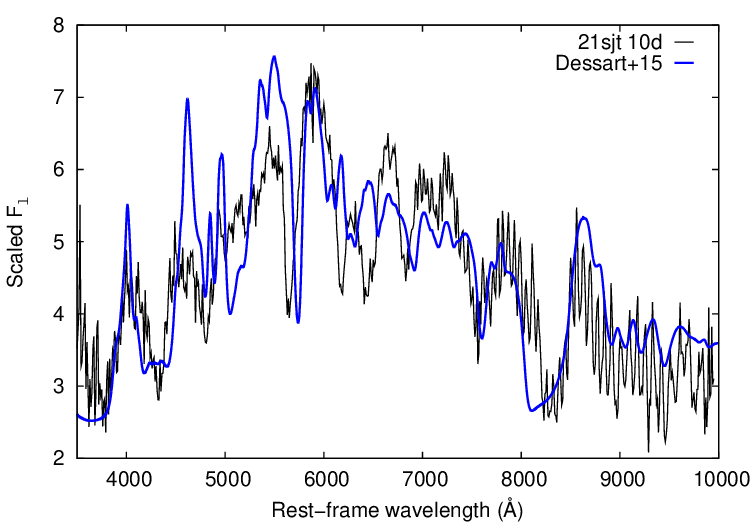}
\includegraphics[width=6cm]{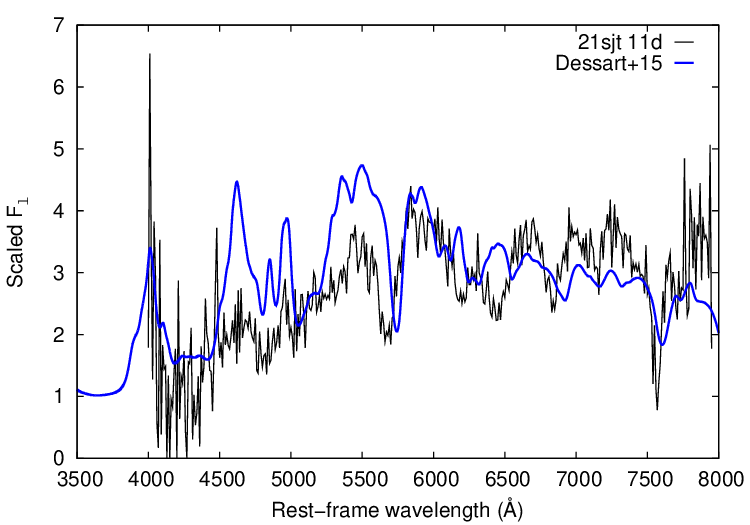}
\includegraphics[width=6cm]{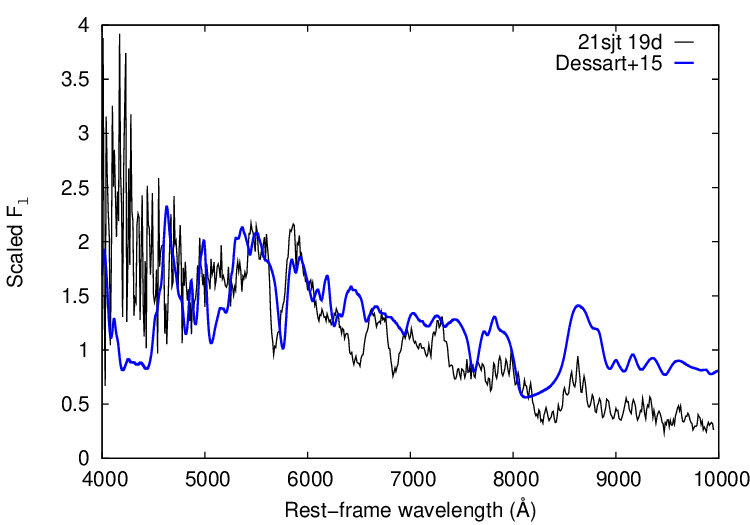}
\caption{The observed spectra of SN~2021sjt corrected for interstellar reddening and redshift (black lines) compared to the synthetic spectra of \citet{2015MNRAS.453.2189D} created for Type Ib SNe (blue lines). Each observed spectrum is compared to a synthetic spectrum with the same phase from the explosion.}.  
\label{fig:dessart}
\end{figure*}

Before modeling, the observed and corrected spectra of SN~2021sjt were compared to the synthetic spectra for Type Ib SNe presented by \citet{2015MNRAS.453.2189D} (see Figure \ref{fig:dessart}) in order to help the classification and the ion identification. As can be seen in Figure \ref{fig:dessart}, SN~2021sjt shows resemblance to the models named 6p5Ax1 built specifically for Type Ib SNe by \citet{2015MNRAS.453.2189D}.

\section{The expanding photosphere method}\label{sec:epm}

The expanding photosphere method (EPM) is a geometric method that is widely applied to calculate the distance of Type IIP SNe  \citep{Dessart_2005, Takats_2012, 2012A&A...540A..93V, Mitchell_2023}, however in the literature it was applied to different types of SNe as well (see e.g. \citet{2004A&A...427..453V,2024AAS...24340702M}. It it based on the comparison of the angular size of the expanding SN shell and the physical radius inferred using the photospheric velocity of the SN \citep{Baade_1926, Wesselink_1946,1974ApJ...193...27K}. Assuming homologous expansion, the angular radius of the photosphere ($\theta$) can be derived as

\begin{equation}
 \theta = \frac{R_{\rm ph}}{D} = {{R_0 + v_{\rm ph} (t-t_0)} \over {D}},
    \label{eq:theta}
\end{equation}

where $R_{\rm ph}$ is the radius of the photosphere, $D$ is the distance of the object, and $R_0$ is the photospheric radius at the moment of the explosion ($t_0$). $R_0$ is usually neglected, since $v_{\rm ph}\cdot\Delta t >> R_0$ from $\sim 5$ days after the moment of the shock breakout \citep{2001ApJ...558..615H, 2012A&A...540A..93V}, thus Eq. \ref{eq:theta} can be simplified and rearranged as
\begin{equation}
 t = D \cdot {{\theta} \over {v_{\rm ph}}} + t_0.
    \label{eq:thetaperve}
\end{equation}
It can be seen from Eq. \ref{eq:thetaperve} that if we estimate the angular radius and the photospheric velocity of the SN measured at different times, the distance and the moment of the explosion can be obtained by linear fitting. In this study, we chose to fix the time of the explosion, since if we have external information on the $t_0$, the final distance estimate can be more precise because of the decreased number of free parameters. This choice is motivated by \citet{2012A&A...540A..93V} as well, who did the same, while determining the distance of M51 using 2 sibling supernovae using the EPM. Thus, before applying Eq. \ref{eq:thetaperve} the time of the explosion was fixed to a date which is between the last non-detection and the first detection of each studied object, and phases were calculated with respect to the used explosion date. For SN~2020dpw, $t_0~=~58905.0$ MJD was used, which is between the discovery date, 2020-02-26 10:01:22.000  (MJD=58905.4) and the last non detection date, 2020-02-25 02:01:22 (MJD=58904.1)  \citet{2020TNSTR.653....1W}. In the case of SN~2021sjt, the time of the explosion was assumed to be 59401.0 MJD, consistently with the moment of discovery 2021-07-07 08:26:52.8 (MJD=2459402.4), and the date of the last non detection 2021-07-05 10:17:46 (MJD=59400.4) \citep{2021TNSTR2352....1F}.  It can be seen that in the case of the studied objects, the difference between the date of the last non-detection and the first detection is less then 1 day, which can be considered as a lucky condition, and gives valid reason to fix $t_0$. The potential systematic error originating from the choice of different $t_0$ values is discussed in Section \ref{err}.

Some assumptions have to be made to estimate $\theta$ from photometry, such as the blackbody radiation of the SN ejecta. However, it is known that photons are usually diluted by electron scattering, thus, a correction factor has to be applied as well that takes into account the dilution from blackbody radiation. This way, the angular radius can be inferred as 
\begin{equation}
    \theta=\sqrt{\frac{F_{\rm bol}}{\xi^2(T)\sigma T^4}},
    \label{eq:theta_final}
\end{equation}
where, $F_{\rm bol}$ is the (pseudo)-bolometric flux, $T$ is the temperature of the photosphere, $\xi(T)$ is the correction factor and $\sigma$ is the Stephan-Boltzmann constant. It is important to note that the precise value of the dilution factor is still under debate, especially for Type Ib SNe \citep[see e.g.][]{1996ApJ...466..911E,Dessart_2005,2015MNRAS.453.2189D}.

Then, the velocity of the photosphere is usually estimated from the Doppler-shift of specific spectral lines, or spectrum modeling. In this paper, we choose to model the overall available spectra in order to obtain the most reliable $v_{\rm ph}$ estimates (see Section \ref{sec:distances}).

Finally, it has to be mentioned that there are some caveats of the EPM, such as i) the assumption of spherical symmetry (e.g. in the case of
SN~2005cs \citet{2007AstL...33..736G}, polarimetric measurements showed a high degree of asymmetry in the ejecta), ii) the sensitivity of the method to the photospheric velocity values, iii) and the debated values of the correction factors ($\xi$). However, its application for SN siblings may open new doors in refining the method, as well as the distances to the host of such SNe.

\section{Distance estimates}\label{sec:distances}

In order to apply the expanding photosphere method, the determination of the angular diameter ($\theta$) and the photospheric velocity ($v_{\rm ph}$) was crucial. The former value was inferred using photometric data, while the latter was obtained by spectrum modeling. After getting $\theta$ and $v_{\rm ph}$ in different epochs, an empirical relation was fitted to the photospheric velocities, and with interpolation, velocities in the time of the photometric measurements were calculated. Finally, the distance and the moment of the explosion were obtained by fitting Eq. 
\ref{eq:thetaperve} to the $\theta / v_{\rm ph}$ values. In the following subsections, the whole procedure is detailed for the specific cases of SN~2020dpw and SN~2021sjt.

\subsection{Angular diameter calculations}

The estimation of $\theta$ was done in the following steps: first, the observed magnitudes were corrected for the interstellar reddening, which was assumed to be $E(B-V)_{\rm tot}$=0.38 (see its explanation in Section \ref{sec:spec}).

After correcting the magnitudes for $E(B-V)$, they were converted to fluxes using the formulae introduced by e.g. \citet{1998A&A...333..231B} and \citet{2016ApJ...822...66F}. In the case of SN~2020dpw, the spectral energy distribution (SED) diagrams were calculated applying the same methodology as e.g. \citet{2019ApJ...870...12L} and \citet{2020ApJ...892..121K} for $BVgriz$ filters, while for SN~2021sjt, $BVgri$ filters were used. It is important to note that $r$ filter covers the place of the $H_\alpha$ line, therefore, to avoid its contamination effects, it was omitted from the next step, where Planck curves were fitted to the SEDs to get the photospheric temperatures ($T_{\rm ph}$) for each observed epoch. 
To calculate the angular radii from Eq. \ref{eq:theta_final}, (pseudo) bolometric flux estimates were necessary as well, which were calculated by integrating the fluxes against wavelength via the trapezoidal rule.

Since neither the UV, nor the IR bands were covered by the used data, they were estimated with extrapolations.
In the UV wavelengths, the flux was presumed to decrease linearly between 2000 \AA\ and $\lambda_B$, and the UV-contribution was estimated from the extinction-corrected $B$-band flux $f_B$ as $f_{bol}^{UV} ~=~ 0.5 f_B (\lambda_B - 2000)$ \citep[see][]{2019ApJ...870...12L}. It is important to note though, that this calculation is tied directly to the $B$-band flux.

In order to get an estimate for the contribution of the unobserved flux in the IR-band, a Rayleigh-Jeans tail was fitted to the corrected $I$- or $z$-band fluxes ($f_I$) and ($f_z$) for SN~2021sjt and SN~2020dpw, respectively, and integrated from $\lambda_I$ or $\lambda_z$ to infinity.  The integration resulted in $f_{bol}^{IR} ~=~ 1.3 f_I \lambda_I / 3$ (or $f_{bol}^{IR} ~=~ 1.3 f_z \lambda_z / 3$), where $f_{bol}^{IR}$ is the contribution of the infrared wavelengths to the overall bolometric flux, which was adopted as the contribution of the missing IR-bands to the bolometric flux. The validity check, as well as the caveats of this method are described in details in \citet{2020ApJ...892..121K}.

To take into account the effect of flux dilution from the black body radiation, the following correction factors ($\xi$) were applied (see Eq. \ref{eq:theta_final})
\begin{equation}
\xi~=~\sum_{i=0}^2{a_i \cdot \left( {10000} \over {T} \right)^i},
    \label{eq:xi}
\end{equation}
where  $a_0~=~0.63241$, $a_1~=~-0.38375$ and $a_2~=~0.28425$ were used in the case of the Type IIp SN~2020dpw \citep{Dessart_2005}. For the Type Ib SN~2021sjt, $a_0~=~1.3178$, $a_1~=~-1.2462$ and $a_2~=~0.6604$ were applied \citep{2015MNRAS.453.2189D}. These constant values are based on B, V, and I-band fluxes.

Finally, the estimated $F_{\rm bol}$, $\xi(T)$ and $T_{\rm ph}$ were substituted into Eq. \ref{eq:theta_final} to derive the angular radii of the two studied SNe in different epochs. Since the EPM can be used reliably only in the case of expanding photosphere, i.e. increasing $\theta$ values, it was applied until the estimated $\theta$ values increased linearly. The inferred angular diameters had a turnover point at 55.63 days and 18.78 days phase post explosion in the case of SN~2020dpw and SN~2021sjt, respectively (see the purple-shaded regions in Figure \ref{fig:lcs}), therefore the EPM was not applied in the later phases, where the $\theta$ was decreasing.

\subsection{Photospheric velocity estimates}

Photospheric velocities of the studied SNe were estimated by modeling the spectra presented in Section \ref{sec:spec}. In the case of SN~2020dpw, only one spectrum taken at 21d after the explosion, had an appropriate signal to noise ratio to identify ions in it with spectrum modeling, while for SN~2021sjt, 5 spectra, observed at 4d, 7d, 10d, 11d and 19d phases were used for velocity calculations. Before the analysis, all spectra were corrected for interstellar reddening and redshift.

Velocity estimates were carried out using the parametrized resonance scattering  code named SYN++ \footnote{\href{https://c3.lbl.gov/es/}{https://c3.lbl.gov/es/}} \citep{2011PASP..123..237T}, which is widely used to model the photospheric phase spectra of supernovae. Using SYN++, some global parameters, such as the photospheric temperature ($T_{\rm ph}$), and the velocity at the photosphere ($v_{\rm ph}$) can be estimated. The contribution of the single ions to the overall model spectra can be taken into account by setting some local parameters interactively and individually. These parameters are the optical depth ($\tau$) of each ion, the minimum and the maximum velocity of the line forming region ($v_{\rm min}$ and $v_{\rm max}$), the scale height of the optical depth above the photosphere (aux), and the excitation temperature ($T_{\rm exc}$). A more detailed description of the model parameters can be read in e.g. \citet{2020ApJ...900...73K,2023ApJ...954...44K}. Similarly to \citet{2022ApJ...940...69K}, the uncertainties of the fitted photospheric velocities were considered to be 1000 km s$^{-1}$.

\begin{figure}[h!]
\centering
\includegraphics[width=8cm]{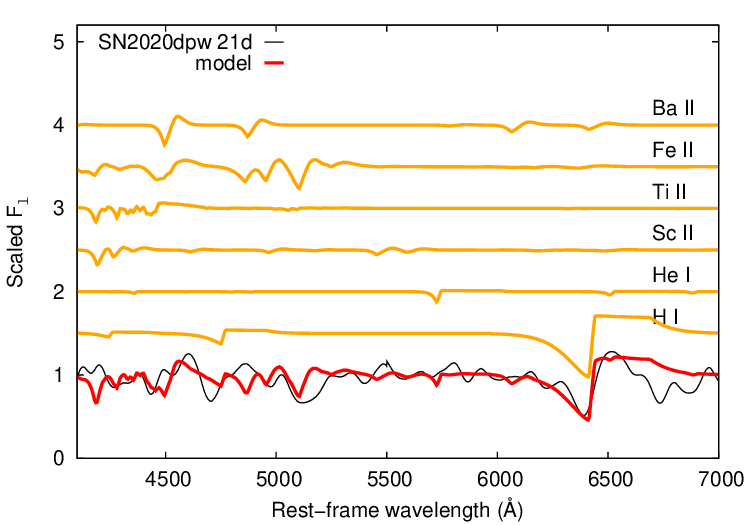}
\caption{The continuum-normalized, redshift- and extinction-corrected spectrum of SN~2020dpw taken at 21d phase post-explosion (black) and its best-fit model obtained in SYN++ (red). The contribution of each identified ion are plotted with orange color, and shifted vertically for clarification.}
\label{fig:20dpw_spec}
\end{figure}

First, the 21d phase spectrum of SN~2020dpw was modeled, where, since this is a Type IIP supernova, the velocity of the photosphere was tied to the Fe II lines. A photospheric temperature of 8000 K and a velocity of 4000 km s$^{-1}$ was found to be the best-fit. In the spectrum, H I, He I, Sc II, Ti II, Fe II and Ba II ions were identified, from which H I, He I and Ti II were considered as high velocity features having a $v_{\rm min}$ of 8000 km s$^{-1}$. Figure \ref{fig:20dpw_spec} shows the redshift- and reddening-corrected, continuum-normalized spectrum of SN~2020dpw together with its best-fit SYN++ model, while Table \ref{tab:local_20dpw} in the Appendix collects the values of the best-fit local parameters.

\begin{figure*}[h!]
\centering
\includegraphics[width=8cm]{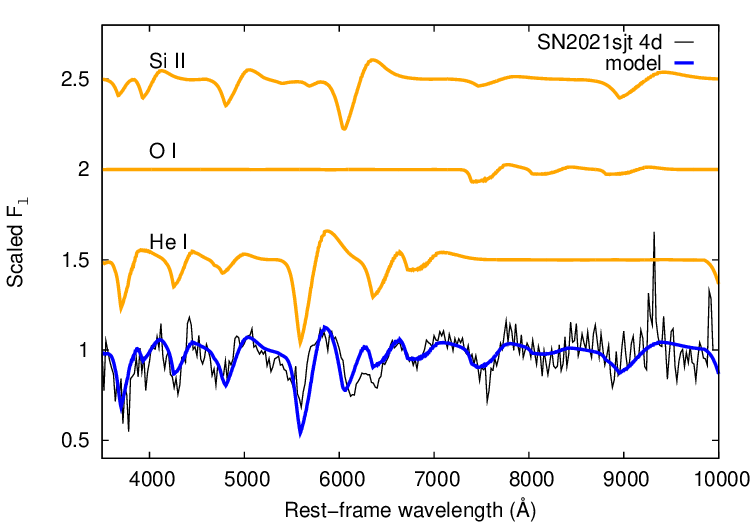}
\includegraphics[width=8cm]{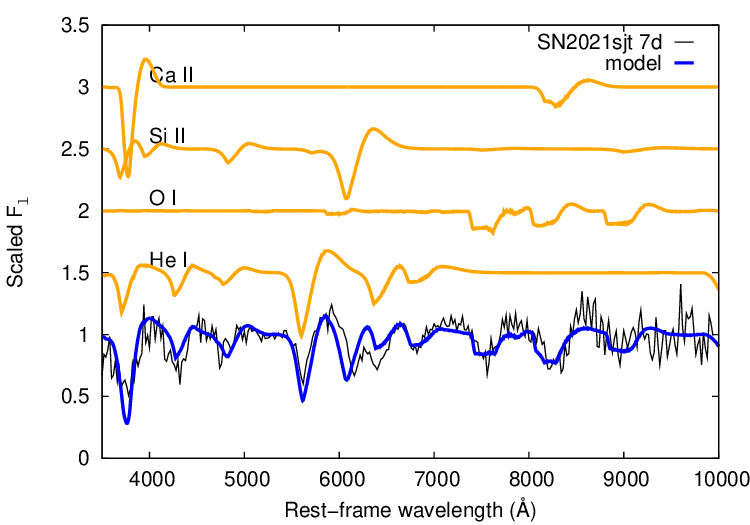}
\includegraphics[width=8cm]{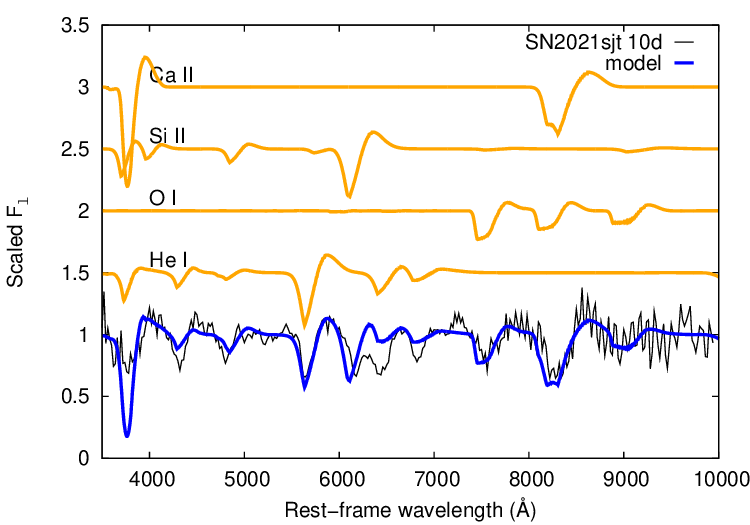}
\includegraphics[width=8cm]{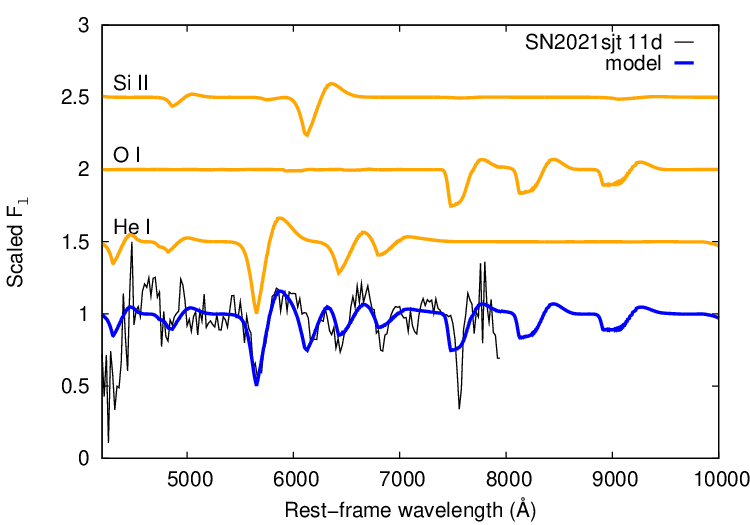}
\includegraphics[width=8cm]{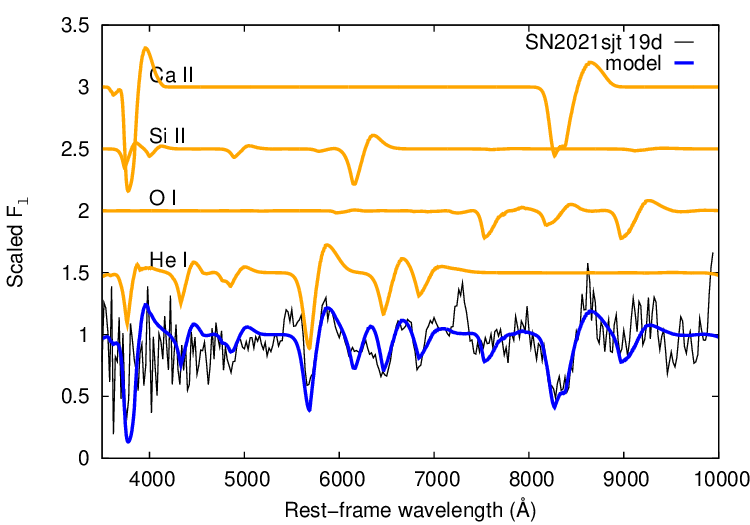}
\caption{The 4d, 7d, 10d, 11d and 19d phase spectrum of SN~2021sjt (black) with their best-fit SYN++ models (blue). The single-ion contribution to the model spectra are plotted in a similar way as in Figure \ref{fig:20dpw_spec}.}
\label{fig:21dpw_spec}
\end{figure*}

Second, the 5 spectra of the Type Ib SN~2021sjt were modeled after the redshift- and reddening corrections. The photospheric temperatures of the
spectrum, taken at 4d phase was found to be 7000 K, which decreased to 5000 K by the last epoch at 19d phase. 
The best-fit $v_{\rm ph}$ at 4d was 15500 km $s^{-1}$, which gradually diminished to 10000 km $s^{-1}$ by 19d after the moment of the explosion. The identified ions were He I, O I, Si II and Ca II in case of all modeled spectra. The best-fit SYN++ models are plotted in Figure \ref{fig:21dpw_spec}, while the values of the best-fit local parameters can be found in Table \ref{tab:local_21sjt} in the Appendix.

\subsection{Fitting empirical formulae to the velocities}

Using the combination of photospheric velocities referring to synthetic spectra, and the line velocities inferred from the Fe II $\lambda$5169 line of a sample of Type IIP supernovae, \citet{Takats_2012} created an empirical formula to describe the photospheric velocity evolution of Type IIP SNe. According to their calculations, the photospheric velocity at a given epoch is proportional to the photospheric velocity 50 days after the explosion as follows:

\begin{equation}
v_{ \rm ph}(t) / v_{\rm ph}(50{\rm d})= {(t/50)^{c} \over \sum_{j=0}^2 b_j(t/50)^j }
\label{eq:vtakats}
\end{equation}
where the values of the coefficients were found to be $b_0=0.467 \pm 0.15$, $b_1=0.327 \pm 0.23$, $b_2=0.174 \pm 0.11$ and $c=-0.210 \pm 0.11$. 

\begin{figure}[h!]
\centering
\includegraphics[width=8cm]{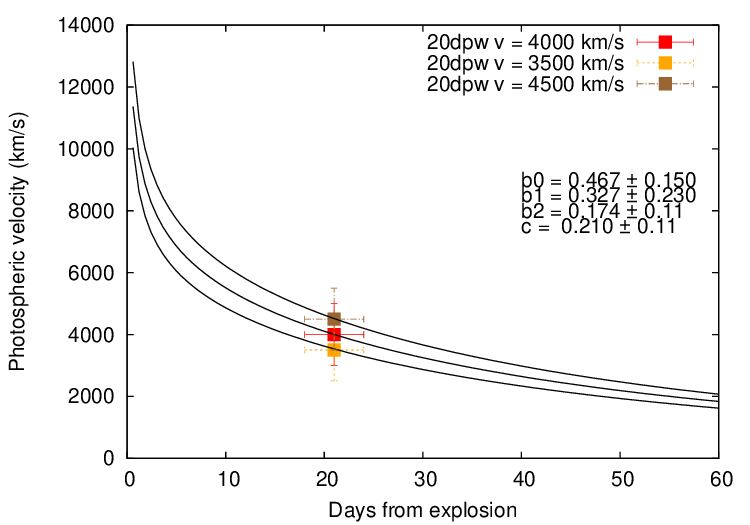}
\caption{Fitting the empirical formula of \citet{Takats_2012}
(Eq. \ref{eq:vtakats}) to the measured photospheric velocity (4000 km s$^{-1}$) of SN~2020dpw (red dot). Orange and brown dots represent a 3500 and a 4500 km s$^{-1}$ test point, which were used to estimate the uncertainty of the $v_{50}$ value.}
\label{fig:velfit_20dpw}
\end{figure}

This formula was fitted to the measured photospheric velocity ($v_{\rm ph}=4000$km s$^{-1}$) of SN~2020dpw (see Figure \ref{fig:velfit_20dpw}), from which the value of the function at +50 days phase ($v_{50}$) was inferred. After that, the same formula was fitted assuming a photospheric velocity of 3500 and 4500 km s$^{-1}$ as well to take into account the uncertainty of the measured $v_{\rm ph}$, and the effect of a different $v_{\rm ph}$ to the estimated $v_{50}$. For $v_{\rm ph}=4000$ km s$^{-1}$, $v_{50}= 2120$ km s$^{-1}$ was obtained, while for $v_{\rm ph}=3500$ km s$^{-1}$, $v_{50}= 1872$ km s$^{-1}$ and for $v_{\rm ph}=4500$ km s$^{-1}$, $v_{50}= 2390$ km s$^{-1}$ was found. The effect of the choice of different $v_{\rm ph}$, and therefore different $v_{50}$ values to the estimated distance is described in Section \ref{dist}, and the uncertainty of the velocities is taken into account in the random error estimates of the $\theta/v$ values.

It is conspicuous that the photospheric velocity of the +21 days phase spectrum of SN~2020dpw is lower compared to the $v_{ph}$ of the Type II-P SNe analyzed by \citet{Takats_2012}. The only exception in their sample is SN~2005cs, the prototypical low luminosity (LL) II-P SN, which was excluded from the fitting in Figure 9 of \citet{Takats_2012}, where the empirical velocity curve that resulted in Eq. \ref{eq:vtakats} was determined. This raises the question: is Eq. \ref{eq:vtakats} applicable to SN~2020dpw at all?

By comparing the bolometric light curve of SN~2020dpw to other well-studied SNe-IIP (see the left panel of Figure \ref{fig:compare} in the Appendix), it can be seen that SN~2020dpw is brighter than SN~2005cs and SN~2020cxd, and has a similar light-curve evolution to SN~1999em. The right panel of Figure \ref{fig:compare} compares the +21d phase spectrum of SN~2020dpw to the similar phase spectra of SN~2005cs and SN~1999em. From that, it is concluded that the photospheric velocity of the spectrum of SN~2020dpw is larger compared to SN~2005cs, and similar to SN~1999em. It is noted that the available spectrum of SN~2020dpw has considerably lower resolution, than the other two SNe, which makes the further comparison difficult. 

If we assume that the velocity evolution of SN~2020dpw is similar to SN~2005cs, it can be seen from the left panel of Figure 9 in \citet{Takats_2012} that the $v_{50}/v_{\sim21}$ ratio of SN~2005cs (marked with purple triangles) is $\sim$2, thus with $v_{21} = 4000$ km s$^{-1}$, the photospheric velocity at 50 days phase is $\sim$ 2000 km s$^{-1}$. This is in agreement with the $v_{50}$ value obtained from the application of Eq. \ref{eq:vtakats} (2120 km s$^-{-1}$) in the case of SN~2020dpw. Therefore it is concluded that the usage of the empirical formula of Eq. \ref{eq:vtakats} can be used in the case of SN~2020dpw in spite of its relatively low photospheric velocity.

Since SN~2021sjt is a Type Ib supernova, the coefficients of Eq. \ref{eq:vtakats} were reconsidered, and given that the evolution of Ib SNe is much faster compared to SNe-IIP, instead of $v_{\rm ph}(50{\rm d}$), $v_{\rm ph}(20{\rm d}$), the photospheric velocity 20 days after the explosion was taken into account.

\begin{figure*}[h!]
\centering
\includegraphics[width=8cm]{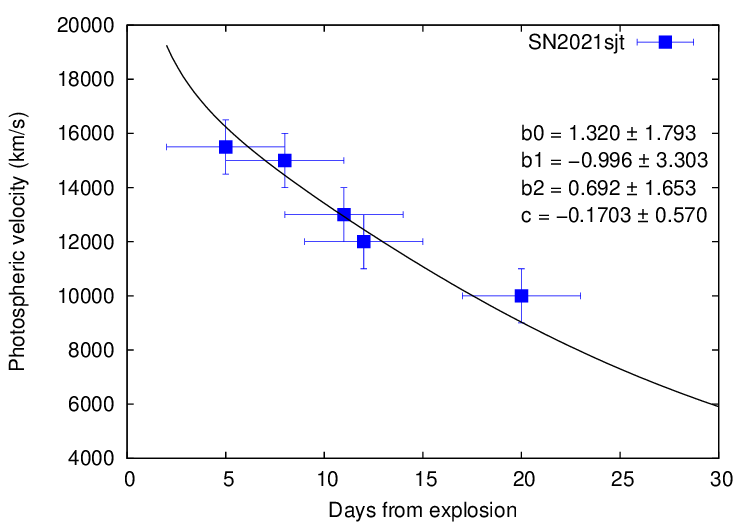}
\caption{Fitting the new empirical formula (Eq. \ref{eq:vuj}) to the estimated $v_{\rm ph}$ values of SN~2021sjt.}
\label{fig:velfit_21sjt}
\end{figure*}

The procedure resulted in a slightly different velocity curve with new coefficient values: 
\begin{equation}
v_{ \rm ph}(t) / v_{\rm ph}(20{\rm d})= {(t/20)^{c} \over \sum_{j=0}^2 b_j(t/20)^j }
\label{eq:vuj}
\end{equation}
where $b_0=1.320 \pm 1,793$, $b_1=-0.996 \pm 3.303$, $b_2=0.629 \pm 1.653$ and $c=-0.170 \pm 0.570$. After getting the new $b_0, b_1$ and $b_2$ values, Eq. \ref{eq:vuj} was fitted to the photospheric velocities of SN~2021sjt estimated with SYN++ modeling (see the right panel of Figure \ref{fig:velfit_21sjt}). It is important to note that the velocities determined in this way have a large uncertainties. 

Having the velocity-curves, it became possible to determine the photospheric velocities in the epochs of the photometric observations (i.e. the inferred angular diameters) of the studied objects with interpolation, and calculate the $\theta/v$ values, which are crucial to determine the distance of the SNe from Eq. \ref{eq:thetaperve}.

\subsection{Distance estimates}\label{dist}

From the inferred $\theta/v$ values, the distance of each object was estimated from Eq. \ref{eq:thetaperve}
with linear fitting. Figure \ref{fig:tavolsag} displays this linear fit for the two studied SNe, while Table \ref{tab:tetaperve} in the Appendix collects the used epochs and the inferred physical parameters, such as SED temperatures, photospheric velocities, angular diameters and $\theta/v$ values for both SN~2020dpw and SN~2021sjt. It is important to note that the results obtained for SN~2021sjt have to be treated cautiously, since the fitting has only a few data points, which are weakly constrained from the available photometric and spectroscopic data.

It can also be seen in Figure \ref{fig:tavolsag} that the data do not fit completely the line in the case of SN~2021sjt, which is because of the fixed explosion time. The optimal fitting to the determined $\theta/v$ values (i.e. when the time of the explosion is not fixed) would result in a $t_0$ contradicting the observed last non-detection and first detection date. Therefore, the we chose to determine $t_0$ from external sources, which helps in making a more relativistic distance estimate. This is also not unexpected that the first data point (and only this one) deviate slightly higher than 1 sigma from the fitted line: it is possible that the basic assumptions of the EPM are not necessarily true for Type Ib SNe in the earliest phases. This example clearly illustrates the key importance of constraining precisely the time of explosion from the last non-detection and the first detection date. To obtain that for a lot of SNe, frequent sampling in the sky surveys will be crucial.

For SN~2020dpw, the fitting resulted in a distance of 25.76$\pm$0.34(random)$\pm$5.51(systematic) MPc, while in the case of SN~2021sjt, D = 24.57$\pm$0.17(random)$\pm$4.67(systematic) Mpc was found. See the detailed description of systematic error estimates in Section \ref{err}.

The distances of the two studied SNe are in agreement within error bars with each other, and they are consistent with the distances calculated for the other SNe exploded in NGC 6951 (see Table \ref{tab:irodalom}.

\begin{figure*}[h!]
\centering
\includegraphics[width=8cm]{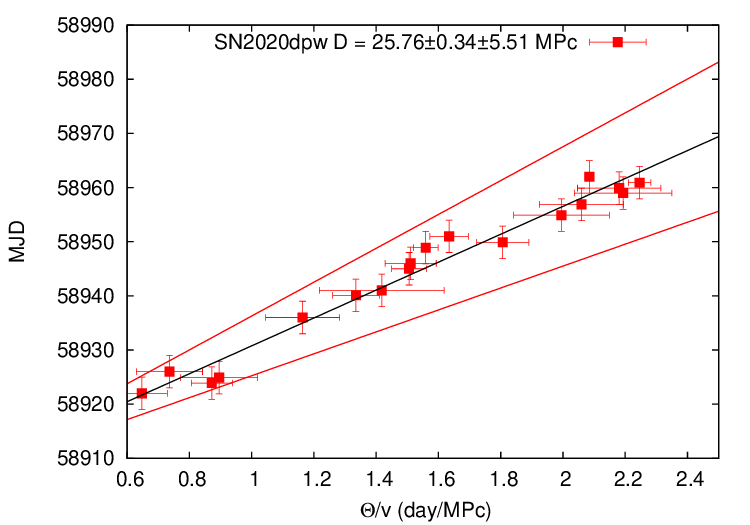}
\includegraphics[width=8cm]{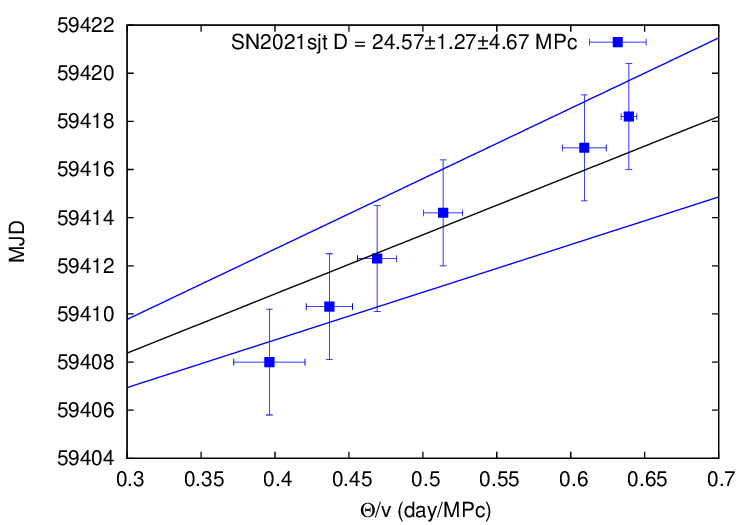}
\caption{Distance determination for SN~2020dpw (left) and SN~2021sjt (right) from Eq. \ref{eq:thetaperve} using the expanding photosphere method. The colored lines correspond to 1 $\sigma$ uncertainty in D, taking into account the systematic errors.
}
\label{fig:tavolsag}
\end{figure*}

\subsection{Error estimates}\label{err}

There are several sources of systematic error in the used EPM. Most importantly, (a) the choice of the $\xi$ values plays critical role in the derived distances. Moreover, among others, (b) the time of the explosion and (c) the interstellar reddening of the host galaxy have effect on the results. Therefore, systematic errors yielded by these three factors were estimated and combined to obtain reliable uncertainty estimates. (See Table \ref{tab:sys} and Figure \ref{fig:sys} for a summarization.)

First, the impact of the \textbf{dilution factor} to the derived distances was tested. In the case of SN~2020dpw, we used the $a_0$, $a_1$ and $a_2$ coefficients for BVI filters published by \citet{Dessart_2005} ($a_0=0.63241, a_1=-0.38375, a_2=0.28425$) to our "best-fit" distance, which are different from the ones obtained by \citet{Eastman_1996} ($a_0=0.686, a_1=-0.577, a_2=0.316$) and \citet{2001ApJ...558..615H} ($a_0=0.7226, a_1=-0.6942, a_2=0.374$) for the same filter combination. Therefore, distances using the latter two correction factors were also calculated, which resulted in a D of 19.75$\pm$0.39 Mpc and 20.76$\pm$0.38 Mpc by using the $\xi$s from the papers of \citet{Eastman_1996} and \citet{2001ApJ...558..615H}, respectively Although this is consistent with \citet{2018A&A...611A..25G}, who showed that distances using the dilution factors of \citet{Dessart_2005} tend to be larger the ones calculated applying the $\xi$ values presented by \citet{2001ApJ...558..615H}, this difference cannot be neglected: it yields a systematic uncertainty of 3.5 MPc. Our choice of the $\xi$ values of \citet{Dessart_2005} are motivated by \citet{2006A&A...447..691D} as well, who measured the distance to SN 1999em and its host using both Cepheids and supernovae, and found that the resulting distances from the two object types are significant different, if one uses the dilution factors of   \citet{2001ApJ...558..615H}, but agree, when  the $\xi$ values published by \citet{Dessart_2005} are applied.

As SN~2021sjt is a Type Ib SN, we used the only $\xi$ values calibrated for this SN type, published by \citet{2015MNRAS.453.2189D}. To estimate systematic error resulting from this choice, we calculated the distance of the object using $\xi~=~1$, which was earlier applied by \citet{Takats_2012}. According to their study, the spectra of Type Ib SNe do not differ significantly from the blackbody radiation, on the contrary to Type IIP SNe, therefore the dilution factor can be approximated with 1. This latter choice yielded a D of 21.9$\pm$1.15 Mpc, which is $\sim$2.5 Mpc lower than the distance using the dilution factors by  \citet{2015MNRAS.453.2189D}. From this, a systematic error of 1.34 Mpc was obtained.

Second, the effect of the choice of the \textbf{fixed $t_0$} was taken into account. In the case of SN~2020dpw, the explosion date was estimated to be 58905.0, a date in between the date of the last non-detection and the first detection. To test the uncertainty caused by fixing this value, new distances were inferred using $t_0$=58904.0 and 58906.0, resulting in $D=26.34 \pm 0.34$ Mpc and $25.18\pm0.34$ Mpc, respectively, yielding a 0.58 Mpc systematic error value.

The same procedure was followed for SN~2021sjt as well: $t_0=59401.0$ was chosen in the first place to be fixed, and here, distances were re-calculated using $t_0=59400.0$ and 59402.0. In the case of the former explosion date value, 26.46$\pm$ 1.12 Mpc, while for the latter, 22.67$\pm$1.41 Mpc was obtained, resulting in a systematic error of 1.9 Mpc.

Third, the effect of the \textbf{reddening} was tested. This is an important factor, since currently, the reddening of the host is unknown, and we only can make assumptions. A value of $E(B-V)=0.38$ was used in the "best-fit" distances, however, there is a possibility that the reddening across the host is not the same for the two studied sibling supernovae. Here, the usage of two additional reddening values was tested: $E(B-V)=0.32$ mag,  if we neglect the reddening of the host galaxy and, and  $E(B-V)=0.44$ mag to take into account an increased reddening. For SN~2020dpw, we obtained $D=26.66 \pm 0.52$ Mpc and $D=23.87 \pm 0.3$ Mpc, respectively, yielding a systematic error of 1.43 Mpc. This uncertainty value was used in the case of SN~2021sjt as well, since the change in the reddening causes the same effect the the flux of the two SNe. It can be seen from the difference between the distances using different $E(B-V)$ values is that the EPM is less sensitive to the interstellar reddening, than the SCM. 

In total, the mentioned factors provided a 5.51 and a 4.67 Mpc uncertainty to the distance estimates of SN~2020dpw and SN~2021sjt, respectively. 

\subsection{Caveats of the velocity estimates}

\begin{figure*}[h!]
\centering
\includegraphics[width=6cm]{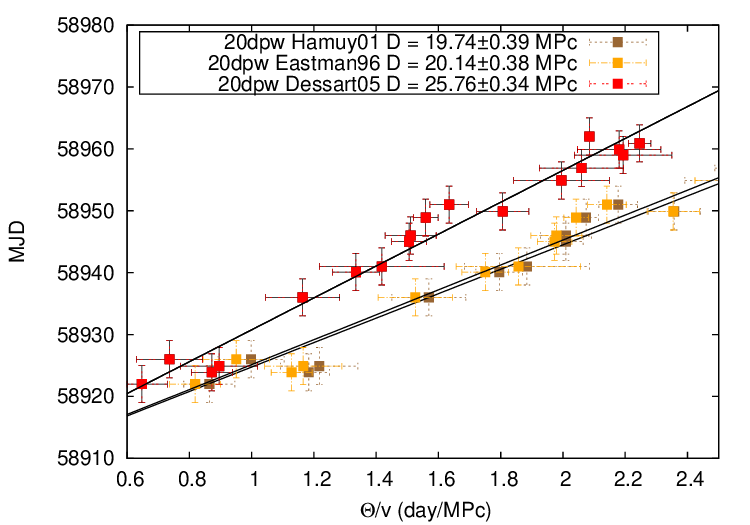}
\includegraphics[width=6cm]{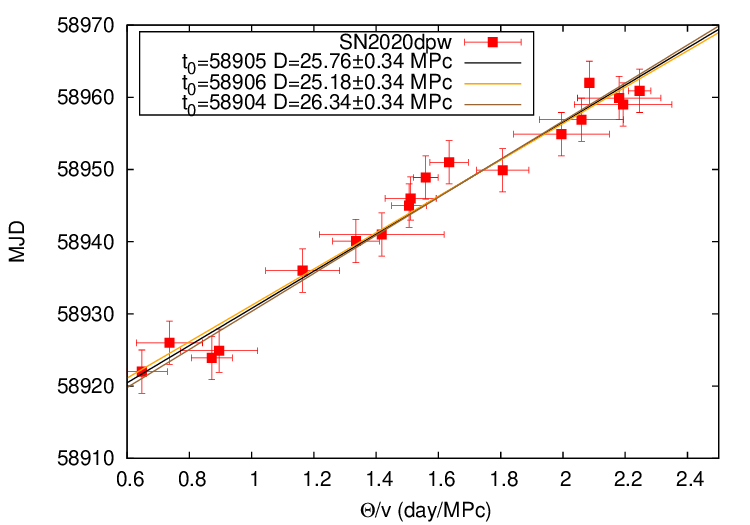}
\includegraphics[width=6cm]{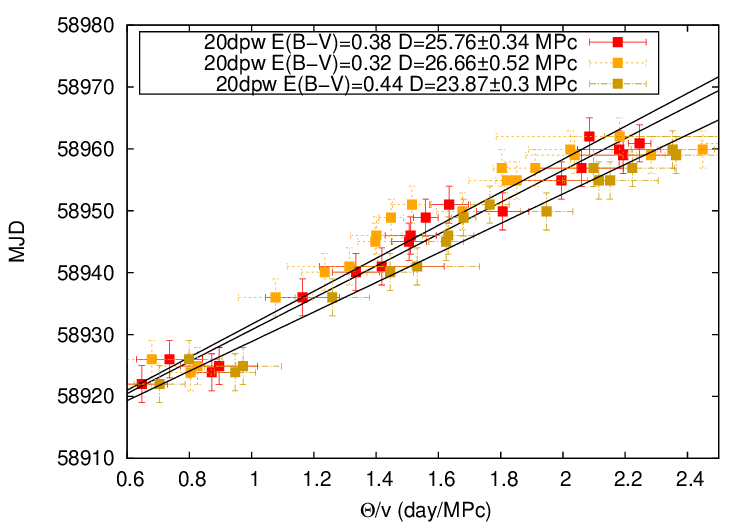}
\includegraphics[width=6cm]{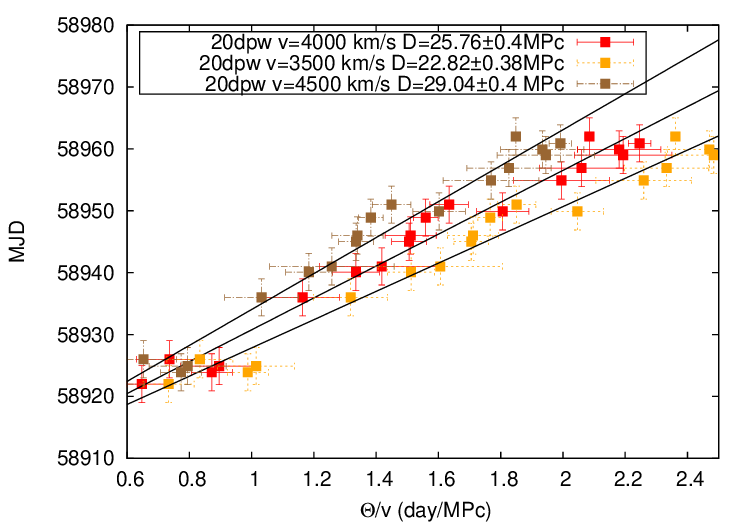}
\includegraphics[width=6cm]{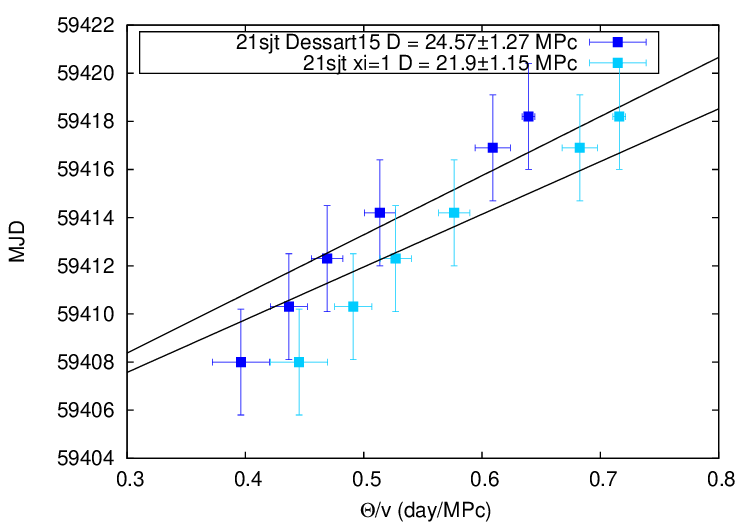}
\includegraphics[width=6cm]{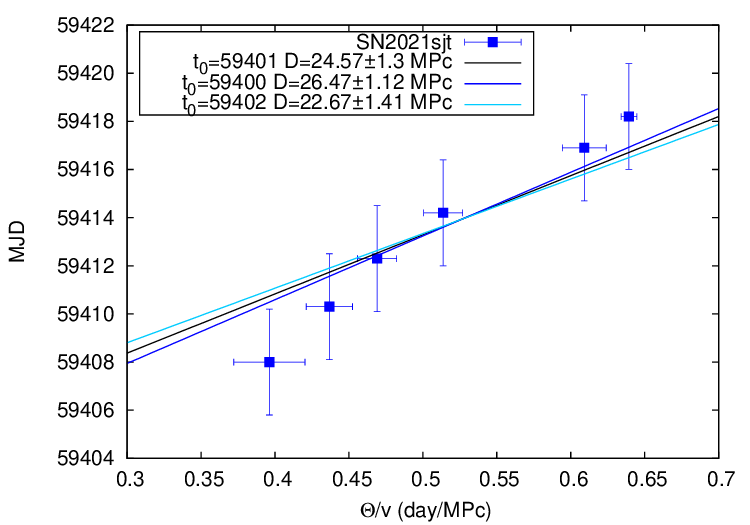}
\caption{Error estimates, and their sources. Top left: distances using different dilution factors, and their effect to the distance of SN~2020dpw. Top middle: the effect of the the change in the fixed $t_0$ to the derived D of SN~2020dpw. Top right: testing the effect of the total interstellar reddening. Bottom left: testing the effect of different $v_{50}$ values for SN~2020dpw. Bottom middle: the same as top left, but for SN~2021sjt. Bottom right: the same as top middle for SN~2021sjt.}
\label{fig:sys}
\end{figure*}

Apart from the systematic errors, the uncertainty of the photospheric velocity in the case of SN~2020dpw, where only one spectrum was available, has to be mentioned. To take this into account, a $+$ and a $-$ 500 km s$^{-1}$ shift in the estimated photospheric velocity were tested. By fitting the empirical formula of Eq. \ref{eq:vtakats} to the different photospheric velocities, 3 different $v_{50}$ were obtained, and used for test distance estimates (see Figure \ref{fig:sys}. The following distances were inferred: 22.82$\pm$0.38 Mpc for $v_{\rm ph}=3500$ km s$^{-1}$, 25.76$\pm$0.4 Mpc for $v_{\rm ph}=4000$ km s$^{-1}$ (the best-fit value) and 29.04$\pm$0.4 Mpc for $v_{\rm ph}=4500$ km s$^{-1}$. The standard deviation of these distances is 3.1 Mpc, which shows that the results strongly depend on the photospheric velocities, and therefore the results have to be treated with caution. 
It is also important to note that the velocity curve introduced in Eq. \ref{eq:vuj} for SN~2021sjt is fitted to only 5 points, which makes the obtained velocities used in the $\theta/v$ calculations questionable. In order to validate the application of Eq. \ref{eq:vuj}, the following test has been made: we interpolated linearly the measured velocities to the dates of the photometric observations, and fitted Eq. \ref{eq:thetaperve} to the newly obtained $\theta/v$ values. This calculation gave a distance of $D=25.19\pm1.47$ Mpc, which is consistent with the "best-fit" distance using the fitted velocity curve.

\begin{table*}
\caption{Error estimates for the studied SNe. Three main factors were taken into account as the sources of systematic error: the usage of different $\xi$ values, the choice of the date of the explosion ($t_0$) and the interstellar reddening ($E(B-V)$). The estimates/approximations used in the case of the "best-fit" distance estimate are marked with boldface.}
\label{tab:sys}
\begin{center}
\begin{tabular}{lc|cc|cc|c}
\hline
\hline
$\xi$ & $D_1$ & $t_0$ & $D_2$ & E(B-V) & $D_3$ & $\sigma_{\rm tot}$ \\
     &  (Mpc)    & (MJD)  & (Mpc)   & (mag)  & (Mpc)   & (Mpc) \\
\hline
SN2020dpw & & & & & & \\
\hline
\citet{2001ApJ...558..615H} & 19.74 $\pm$ 0.39 &  58904.0 & 26.34 $\pm$0.34 &0.32 & 26.66 $\pm$0.52 & \\ 
 \citet{1996ApJ...466..911E} & 20.14 $\pm$ 0.38 & \bf{58905.0} & 25.76 $\pm$0.34  & {\bf 0.38} & 25.76 $\pm$0.34 & \\
\bf{  \citet{Dessart_2005}} & 25.76 $\pm$ 0.34 & 58906.0 & 25.18 $\pm$0.34  & 0.44 & 23.87 $\pm$ 0.3 & \\
 \hline
 $\sigma_{\rm sys},_\xi = $ & 3.5 & $\sigma_{\rm sys},_{t_0} = $ & 0.58 & $\sigma_{\rm sys},_{\rm ebv} =$ & 1.43 & 5.51 \\
 \hline
SN2021sjt &&&&&& \\
\hline 
\bf \citet{2015MNRAS.453.2189D}{} & 24.57 $\pm$ 1.27 & 59400.0 & 26.47 $\pm$ 1.12 & & & \\
1 \citep{Takats_2012} & 21.9 $\pm$ 1.15 & \bf{59401.0} & 24.57 $\pm$1.30 & & & \\
                       &                 & 49402.0 & 22.67 $\pm$1.41 & & & \\
\hline 
 $\sigma_{\rm sys},_\xi = $ & 1.34 & $\sigma_{\rm sys},_{t_0} = $ & 1.90 & $\sigma_{\rm sys},_{\rm ebv} =$ & 1.43 & 4.67 \\
 \hline
\end{tabular}
\end{center}
\end{table*}

\section{Conclusions}\label{sec:conc}

In this paper, we presented new distance estimates to two sibling SNe exploded in NGC6951: the Type IIP SN~2020dpw and the Type Ib SN~2021sjt. SN siblings are powerful tools to refine extragalactic distance measurement techniques, and as such, they can be used to obtain reliable distance estimates to their host galaxies. Our study aims to give new estimates on the distance of NGC6951, and to test the usability of the expanding photosphere method on not only Type IIP SNe, but Type Ib SNe as well.

To infer the distance to the two studied SNe, we applied the expanding photosphere method, for which the estimation of the angular diameter ($\theta$) and the photospheric velocity ($v_{\rm ph}$) were required. The former was derived from publicly available and new photometric data from the RC80, the BRC80, the LCO 1 and 2 m telescopes. In order to obtain $\theta$, pseudo-bolometric flux and photospheric temperature calculations were carried out. In the case of SN~2020dpw the SEDs for different observational epochs were generated from $BVgriz$ data, while for SN~2021sjt, $BVgri$ data were used. The SEDs were fitted with blackbody curves to estimate the photospheric temperatures, for which, a correction factor was applied to account for the dilution from the blackbody radiation due to electron scattering. The pseudo-bolometric fluxes were inferred using the trapezoidal rule. From these quantities, estimates for $\theta$ were made.

Next, the photospheric velocities of the studied SNe were done by SYN++ spectrum modeling. In the case of SN~2020dpw, only one, while for SN~2021sjt, 5 spectra were available. To obtain $v_{\rm ph}$ in the epoch of the photometric observations, the empirical formula of \citet{Takats_2012} was applied in the case of the Type IIP SN~2020dpw. This formula was altered to get the velocity curve of SN~2021sjt. This way, it became possible to derive photospheric velocity values at the same epochs as the photospheric observations. 

From the obtained $\theta$ and $v_{\rm ph}$ in different epochs, the distance was estimated via linear fitting for both SNe. The time of the explosion was fixed to a date between the last non-detection and the first detection. The distance of SN~2020dpw was found to be  $25.76 \pm 0.34(\rm random) \pm 5.51$(systematic) Mpc, while for SN~2021sjt our calculations using the preferred dilution factor, time of explosion and interstellar reddening resulted in D~=~$24.57 \pm 1.27 (\rm random)\pm 4.67$(systematic) Mpc. These values are roughly consistent with the literature and with each other. Systematic errors originating from the choice of the dilution factor, the time of explosion and interstellar reddening were taken into account.

Our study showed that the expanding photosphere method can be applied not only to Type IIP, but to Type Ib SNe as well, and
our analysis has demonstrated the reason d'etre of the method. The distance estimate to other galaxies that host two or more different type of supernovae is planned in future studies. This way, the distance estimate techniques can be tested and calibrated, and the distance estimates to specific galaxies and objects can be finetuned.

\begin{acknowledgements}

This work makes use of observations from the Las Cumbres Observatory network.
The LCO team is supported by NSF grants AST-2308113 and AST-1911151. J.R.F. is supported by the U.S. National
Science Foundation (NSF) Graduate Research Fellowship Program under grant No. 2139319. V.K. is supported by the Hungarian National Research, Development, and Innovation Office grant KKP-143986. 
A.P. and R.Sz. are supported by the Hungarian National Research, Development, and Innovation Office (NKFIH) K-138962 grant.
S.V.  acknowledge support by NSF grants AST-2407565.  We are indebted to the anonymous referee, who helped with constructive criticism and suggestions to improve the quality of the paper.

\end{acknowledgements}

%
%

\bibliographystyle{aa}
\bibliography{main}

\begin{appendix} 

\section{Supplementary material}

\begin{table*}
\caption{Photometric data of SN~2020dpw taken with RC80.}
\label{tab:20dpw_lc} 
\begin{center}
\begin{tabular}{lccccccc}
\hline
\hline
MJD & B & V & g & r & i & z \\
\hline
58922.03 & 17.853 (0.090) & 17.053 (0.039) & 17.294 (0.039) & 16.813 (0.035) & 16.757 (0.051) & 16.683 (0.093) \\
58923.91 & 17.845 (0.091) & 16.958 (0.042) & 17.252 (0.043) & 16.729 (0.035) & 16.558 (0.054) & 16.444 (0.077) \\
58924.92 & 17.996 (0.082) & 17.145 (0.047) & 17.429 (0.049) & 16.817 (0.035) & 16.671 (0.054) & 16.529 (0.075) \\
58925.97 & 18.376 (0.098) & 17.357 (0.047) & 17.640 (0.045) & 17.112 (0.048) & 16.979 (0.070) & 16.827 (0.079) \\
58936.04 & 18.614 (0.096) & 17.308 (0.038) & 17.759 (0.041) & 17.016 (0.038) & 16.845 (0.051) & 16.687 (0.069) \\
58940.05 & 18.444 (0.101) & 17.315 (0.047) & 17.757 (0.050) & 16.961 (0.042) & 16.782 (0.058) & 16.693 (0.084) \\
58940.97 & 18.707 (0.128) & 17.245 (0.048) & 17.820 (0.064) & 16.983 (0.043) & 16.769 (0.054) & 16.575 (0.075) \\
58944.97 & 18.834 (0.175) & 17.328 (0.057) & 17.784 (0.076) & 16.932 (0.045) & 16.729 (0.054) & 16.639 (0.083) \\
58946.01 & 18.806 (0.151) & 17.379 (0.052) & 17.866 (0.070) & 17.048 (0.042) & 16.814 (0.053) & 16.641 (0.074) \\
58948.94 & 19.101 (0.182) & 17.426 (0.064) & 17.882 (0.071) & 17.002 (0.045) & 16.828 (0.055) & 16.663 (0.074) \\
58949.94 & 18.811 (0.182) & 17.364 (0.085) & 17.902 (0.062) & 16.945 (0.043) & 16.661 (0.050) & 16.588 (0.081) \\
58950.99 & 18.816 (0.117) & 17.410 (0.047) & 17.879 (0.049) & 17.078 (0.041) & 16.831 (0.055) & 16.687 (0.071) \\
58954.93 & 18.938 (0.102) & 17.464 (0.043) & 17.895 (0.041) & 17.023 (0.037) & 16.786 (0.050) & 16.558 (0.063) \\
58956.95 & 18.884 (0.131) & 17.460 (0.048) & 17.921 (0.053) & 17.072 (0.041) & 16.819 (0.058) & 16.479 (0.112) \\
58958.96 & 19.035 (0.130) & 17.463 (0.050) & 17.896 (0.054) & 16.989 (0.045) & 16.758 (0.061) & 16.479 (0.080) \\
58959.95 & 18.876 (0.122) & 17.441 (0.051) & 17.969 (0.059) & 17.003 (0.046) & 16.787 (0.060) & 16.667 (0.085) \\
58954.93 & 18.938 (0.102) & 17.464 (0.043) & 17.895 (0.041) & 17.023 (0.037) & 16.786 (0.050) & 16.558 (0.063) \\
58956.95 & 18.884 (0.131) & 17.460 (0.048) & 17.921 (0.053) & 17.072 (0.041) & 16.819 (0.058) & 16.479 (0.112) \\
58958.96 & 19.035 (0.130) & 17.463 (0.050) & 17.896 (0.054) & 16.989 (0.045) & 16.758 (0.061) & 16.479 (0.080) \\
58959.95 & 18.876 (0.122) & 17.441 (0.051) & 17.969 (0.059) & 17.003 (0.046) & 16.787 (0.060) & 16.667 (0.085) \\
58960.91 & 18.851 (0.128) & 17.474 (0.055) & 17.997 (0.058) & 16.942 (0.044) & 16.683 (0.055) & 16.553 (0.085) \\
58962.03 & 19.134 (0.108) & 17.528 (0.051) & 17.993 (0.052) & 17.047 (0.043) & 16.806 (0.055) & 16.614 (0.071) \\
59028.99 & 21.962 (0.073) & 19.186 (0.178) & 19.918 (0.226) & 18.490 (0.145) & 18.047 (0.152) & 17.563 (0.173) \\
59060.93 & $-$ ($-$)& 19.985 (0.595) & 22.145 (0.034) & 18.725 (0.203) & 18.532 (0.303) & 17.697 (0.219) \\
59061.87 & $-$ ($-$)& 19.462 (0.319) & 20.356 (0.649) & 18.888 (0.272) & 18.116 (0.220) & 17.439 (0.166) \\
59062.93 & 21.198 (0.094) & 19.758 (0.792) & 19.765 (0.552) & 18.871 (0.369) & 18.562 (0.579) & 17.880 (0.469) \\
59068.98 & 20.333 (0.853) & 19.426 (0.639) & 19.914 (0.710) & 18.609 (0.232) & 18.542 (0.347) & 18.112 (0.427) \\
59069.95 & $-$ ($-$)& 19.582 (0.466) & 21.599 (0.037) & 18.710 (0.191) & 18.400 (0.243) & 17.511 (0.164) \\
59070.90 & 22.301 (0.087) & 19.358 (0.310) & 22.473 (0.038) & 18.623 (0.188) & 18.324 (0.243) & 17.858 (0.270) \\
59071.89 & 20.546 (1.982) & 19.663 (0.934) & 20.497 (2.066) & 18.538 (0.208) & 18.326 (0.256) & 17.484 (0.222) \\
59074.02 & 20.488 (0.528) & 19.632 (0.323) & 20.782 (0.802) & 18.694 (0.180) & 18.417 (0.264) & 17.649 (0.207) \\
59074.92 & 21.721 (2.385) & 19.633 (0.312) & 20.389 (0.494) & 18.629 (0.143) & 18.333 (0.264) & 17.841 (0.274) \\
59075.92 & 23.037 (0.081) & 19.620 (0.359) & 20.665 (0.685) & 18.712 (0.175) & 18.310 (0.204) & 18.066 (0.323) \\
59083.05 & 21.865 (1.676) & 19.728 (0.297) & 20.360 (0.323) & 18.767 (0.167) & 18.610 (0.255) & 18.007 (0.238) \\
59083.88 & 20.780 (0.342) & 19.503 (0.222) & 20.153 (0.271) & 18.552 (0.135) & 18.334 (0.198) & 17.631 (0.193) \\
59084.90 & 21.226 (0.940) & 19.224 (0.177) & 20.356 (0.420) & 18.747 (0.166) & 18.373 (0.212) & 17.779 (0.207) \\
59086.90 & 22.106 (2.916) & 19.592 (0.276) & 20.308 (0.341) & 18.765 (0.161) & 18.602 (0.272) & 17.705 (0.197) \\
59088.07 & 22.610 (0.089) & 20.607 (1.455) & 21.197 (1.676) & 18.814 (0.209) & 18.637 (0.307) & 17.748 (0.207) \\
59089.08 & 21.267 (1.444) & 19.921 (0.441) & 20.438 (0.434) & 18.940 (0.216) & 18.711 (0.298) & 18.004 (0.226) \\
59101.02 & $-$ ($-$)& 20.359 (0.618) & 20.259 (0.343) & 19.004 (0.231) & 19.075 (0.405) & 18.195 (0.272) \\
\hline
\end{tabular}
\end{center}
\end{table*}

\begin{table*}
\caption{Photometric data of SN~2021sjt taken with BRC80.}
\label{tab:21sjt_lc_brc} 
\begin{center}
\begin{tabular}{lccccccc}
\hline
\hline
MJD & B & V & g & r & i & z \\
\hline
59415.95 & 19.331 (0.083) & 17.920 (0.045) & 18.400 (0.034) & 17.176 (0.032) & 16.528 (0.029) & 16.321 (0.058) \\
59446.91 & 20.468 (0.166) & 18.970 (0.077) & 19.599 (0.088) & 18.011 (0.043) & 17.411 (0.040) & 16.851 (0.054) \\
59529.69 & 20.843 (0.188) & 19.754 (0.137) & 20.256 (0.165) & 18.511 (0.060) & 18.005 (0.055) & 17.815 (0.097) \\
59511.03 & 19.915 (0.188) & 18.658 (0.104) & 19.217 (0.122) & 17.836 (0.077) & 17.428 (0.085) & 17.014 (0.089) \\
59447.98 & 19.723 (0.130) & 19.100 (0.092) & 19.725 (0.123) & 18.322 (0.071) & 17.485 (0.044) & 16.864 (0.057) \\
59533.75 & 20.428 (0.221) & 19.266 (0.123) & 19.632 (0.136) & 18.513 (0.081) & 17.977 (0.082) & 18.097 (0.143) \\
59468.82 & 21.112 (0.190) & 18.962 (0.072) & 19.581 (0.079) & 18.243 (0.042) & 17.589 (0.037) & 17.049 (0.062) \\
59418.92 & 19.689 (0.171) & 17.957 (0.056) & 18.613 (0.073) & 17.260 (0.050) & 16.482 (0.042) & 16.229 (0.055) \\
59446.00 & 20.064 (0.108) & 18.828 (0.079) & 19.260 (0.083) & 17.992 (0.054) & 17.303 (0.046) & 16.628 (0.061) \\
59407.98 & 19.337 (0.092) & 17.829 (0.051) & 18.329 (0.047) & 17.232 (0.050) & 16.748 (0.042) & 16.726 (0.071) \\
59517.96 & 19.910 (0.278) & 18.845 (0.272) & 19.691 (0.290) & 18.062 (0.245) & 17.201 (0.255) & 16.921 (0.276) \\
59424.88 & 20.205 (0.129) & 18.260 (0.057) & 18.916 (0.069) & 17.476 (0.044) & 16.744 (0.038) & 16.339 (0.069) \\
59444.98 & 20.553 (0.138) & 18.822 (0.055) & 19.419 (0.067) & 18.037 (0.032) & 17.319 (0.038) & 16.750 (0.048) \\
59465.87 & 20.398 (0.106) & 18.954 (0.058) & 19.761 (0.087) & 18.284 (0.039) & 17.566 (0.032) & 17.288 (0.067) \\
59460.86 & 20.277 (0.106) & 19.095 (0.076) & 19.821 (0.086) & 18.297 (0.039) & 17.526 (0.030) & 17.040 (0.061) \\
59472.84 & 20.507 (0.335) & 19.447 (0.147) & 19.925 (0.148) & 18.303 (0.053) & 17.543 (0.046) & 17.431 (0.105) \\
59432.93 & 20.117 (0.113) & 18.522 (0.058) & 19.206 (0.061) & 17.830 (0.041) & 17.131 (0.035) & 16.716 (0.061) \\
59439.86 & 19.975 (0.092) & 18.831 (0.071) & 19.362 (0.073) & 17.915 (0.041) & 17.230 (0.033) & 16.664 (0.061) \\
59481.81 & 20.601 (0.149) & 19.238 (0.090) & 19.741 (0.092) & 18.368 (0.046) & 17.699 (0.041) & 17.388 (0.069) \\
59451.94 & 20.086 (0.152) & 18.737 (0.074) & 20.172 (0.197) & 18.080 (0.056) & 17.422 (0.065) & 16.706 (0.063) \\
59505.91 & 22.695 (2.797) & $-$ ($-$)& 19.263 (0.149) & 18.519 (0.107) & 17.748 (0.099) & 17.283 (0.144) \\
59470.07 & 20.232 (0.147) & 18.810 (0.096) & 19.420 (0.106) & 18.205 (0.087) & 17.426 (0.099) & 17.015 (0.117) \\
59488.91 & 20.624 (0.178) & 19.137 (0.117) & 19.648 (0.108) & 18.359 (0.079) & 17.608 (0.076) & 17.094 (0.093) \\
59425.99 & 19.885 (0.120) & 18.696 (0.059) & 19.041 (0.055) & 17.604 (0.030) & 16.871 (0.043) & 16.461 (0.043) \\
59437.87 & 20.422 (0.126) & 18.672 (0.065) & 19.144 (0.060) & 17.847 (0.052) & 17.271 (0.045) & 16.653 (0.064) \\
59433.89 & 21.459 (0.295) & 18.661 (0.067) & 19.249 (0.056) & 17.883 (0.044) & 17.174 (0.049) & 16.762 (0.061) \\
59604.19 & 21.671 (0.653) & 19.655 (0.166) & 19.860 (0.127) & 18.502 (0.081) & 17.861 (0.071) & 17.647 (0.103) \\
59416.88 & 19.259 (0.103) & 17.911 (0.051) & 18.449 (0.055) & 17.212 (0.038) & 16.453 (0.034) & 16.300 (0.062) \\
59459.98 & 21.643 (0.296) & 19.078 (0.098) & 19.753 (0.096) & 18.184 (0.053) & 17.520 (0.043) & 16.939 (0.049) \\
59427.07 & 20.739 (0.289) & 18.369 (0.085) & 19.198 (0.088) & 17.656 (0.052) & 16.857 (0.066) & 16.510 (0.082) \\
59491.83 & 20.288 (0.107) & 19.192 (0.073) & 19.658 (0.098) & 18.329 (0.048) & 17.697 (0.038) & 17.317 (0.058) \\
\hline
\end{tabular}
\end{center}
\end{table*}

\begin{table*}
\caption{Photometric data of SN~2021sjt taken with the instruments of LCO.}
\label{tab:21sjt_lc_lco} 
\begin{center}
\begin{tabular}{lcccccc}
\hline
\hline
MJD & B & V & g & r & i \\
\hline
59410.35 & 19.158 (0.066) & 17.728 (0.049) & 18.299 (0.042) & 17.128 (0.043) & 16.586 (0.043) \\
59412.31 & 19.357 (0.065) & 17.757 (0.037) & 18.358 (0.041) & 17.177 (0.035) & 16.576 (0.034) \\
59414.17 & 19.309 (0.064) & 17.757 (0.034) & 18.370 (0.044) & 17.172 (0.026) & 16.544 (0.029) \\
59414.98 & 19.370 (0.063) & 17.789 (0.031) & $-$ ($-$)& 17.252 (0.022) & 16.608 (0.020) \\
59418.19 & 19.480 (0.079) & 17.738 (0.039) & 18.468 (0.043) & 17.122 (0.031) & 16.404 (0.031) \\
59421.20 & 19.801 (0.104) & 17.957 (0.051) & 18.685 (0.066) & 17.251 (0.038) & 16.528 (0.027) \\
59424.14 & 19.887 (0.082) & 18.158 (0.037) & 18.982 (0.047) & 17.490 (0.027) & 16.703 (0.027) \\
59430.17 & 19.990 (0.073) & 18.350 (0.050) & 18.957 (0.063) & 17.608 (0.046) & 16.866 (0.046) \\
59443.90 & 20.403 (0.108) & 18.734 (0.059) & 19.354 (0.062) & 18.067 (0.037) & 17.309 (0.042) \\
59450.09 & 20.483 (0.116) & 18.760 (0.055) & 19.387 (0.069) & 18.167 (0.040) & 17.431 (0.036) \\
59457.08 & $-$ ($-$)& 18.775 (0.054) & $-$ ($-$)& 18.166 (0.050) & 17.390 (0.050) \\
59462.33 & $-$ ($-$)& 18.736 (0.062) & $-$ ($-$)& 18.144 (0.058) & 17.406 (0.057) \\
59467.33 & $-$ ($-$)& 18.709 (0.067) & $-$ ($-$)& 18.149 (0.063) & 17.432 (0.059) \\
59472.32 & $-$ ($-$)& 18.692 (0.068) & $-$ ($-$)& 18.163 (0.065) & 17.468 (0.060) \\
59480.27 & $-$ ($-$)& 18.925 (0.079) & $-$ ($-$)& 18.389 (0.056) & 17.661 (0.089) \\
59487.99 & $-$ ($-$)& 18.992 (0.061) & $-$ ($-$)& 18.575 (0.136) & 17.738 (0.094) \\
59496.24 & $-$ ($-$)& 18.791 (0.074) & $-$ ($-$)& 18.223 (0.074) & 17.235 (0.076) \\
59504.18 & $-$ ($-$)& 18.919 (0.075) & $-$ ($-$)& 18.406 (0.060) & 17.552 (0.065) \\
59511.91 & $-$ ($-$)& 19.022 (0.063) & $-$ ($-$)& 11.600 (2.146) & 11.379 (1.984) \\
59519.87 & $-$ ($-$)& 18.940 (0.069) & $-$ ($-$)& 12.841 (2.133) & 17.640 (0.070) \\
59528.06 & $-$ ($-$)& 18.818 (0.079) & $-$ ($-$)& 18.366 (0.072) & 17.670 (0.067) \\
59536.07 & $-$ ($-$)& 18.696 (0.098) & $-$ ($-$)& 17.990 (0.076) & 17.319 (0.089) \\
\hline
\end{tabular}
\end{center}
\end{table*}

\begin{table*}
\caption{Log of the spectroscopic observations for SN~2020dpw and SN~2021sjt.}
\label{tab:spectra} 
\begin{center}
\begin{tabular}{lccccc}
\hline
\hline
Instrument & Observation date & MJD & Phase & Range & R  \\
            &                 &      & (days) & (\AA) & ($\lambda / \Delta\lambda$) \\
\hline 
 && SN~2020dpw  & & \\ 
ALPY600 & 2020-03-18 & 58926 & 20 & 4000 $-$ 7300  & 100\\
\hline 
 && SN~2021sjt && \\
OGG 2m FLOYDS & 2021-07-10 & 59405 & 13 & 3500 $-$ 10000  & 400-700 \\
OGG 2m FLOYDS & 2021-07-13 & 59408 &15  & 3500 $-$ 10000  & 400-700 \\
OGG 2m FLOYDS & 2021-07-16 & 59411 &19  & 3500 $-$ 10000   & 400-700 \\
OGG 2m FLOYDS & 2021-07-17 & 59412 &20  & 3500 $-$ 10000  & 400-700 \\
OGG 2m FLOYDS & 2021-07-25 & 59420 &28  & 3500 $-$ 10000  & 400-700 \\
\hline
\end{tabular}
\end{center}
\end{table*}

\newpage

\begin{figure*}[h!]
\centering
\includegraphics[width=8cm]{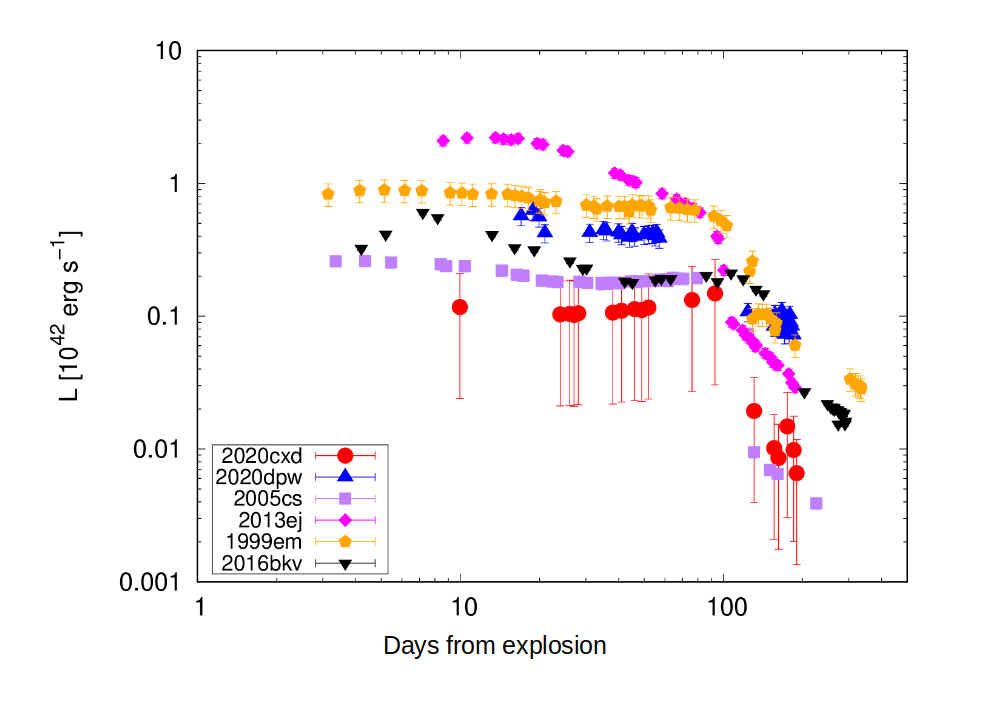}
\includegraphics[width=8cm]{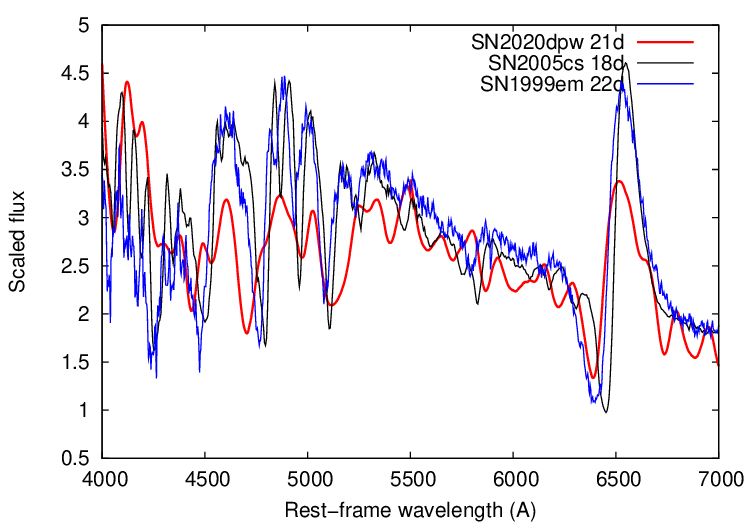}
\caption{Left panel: the comparison of the bolometric light curve of SN2020dpw (blue triangles) with the LCs other Type II-P and low luminosity Type II-P SNe. It can be seen that SN2020dpw is more luminous, than the typical LL II-P SN2005cs, and SN2020cxd, and similar to SN1999em in its light-curve evolution. Right panel: the comparison of similar phase spectra of SN2020dpw (red line), SN2005cs (black line) and SN1999em (blue line). }
\label{fig:compare}
\end{figure*}

\begin{table*}
\caption{Best-fit {\tt SYN++} parameter values of the modeled, 21d phase spectrum of SN~2020dpw. The global parameters are $T_{\rm ph}$ (1000K) and $v_{\rm ph}$ ($10^3$ km s$^{-1}$), while the local parameters are: $\log\tau$ (--),  $v_{\rm min}$ ($10^3$ km s$^{-1}$), $v_{\rm max}$ ($10^3$ km s$^{-1}$), aux ($10^3$ km s$^{-1}$) and $T_{\rm exc}$ (1000 K).
}
\label{tab:local_20dpw} 
\begin{center}
\begin{tabular}{lcccccc}
\hline
\hline
Ions & H I &  He I  & Sc II & Ti II & Fe II & Ba II \\
\hline
  \multicolumn{6}{c}{SN~2020dpw 21d phase ($T_{\rm ph}~=~$8; $v_{\rm ph}~=~$4) } \\
\hline
$\log\tau$ & 0.5 & 0.8  & -0.8&1.0 & 0.0 &-0.2 \\
$v_{\rm min}$ & 7.0 & 8.0  & 4.0&8.0 & 4.0 &4.0\\
$v_{\rm max}$ & 30.0 & 30.0  &30.0 &30.0 & 30.0 & 30.0\\
aux & 4.0 & 1.0 &  1.0&1.0& 2.0 &1.0 \\
$T_{\rm exc}$ &5.0 & 8.0  &8.0& 8.0 &8.0 &8.5\\
\hline
\end{tabular}
\end{center}
\end{table*}

\begin{table*}
\caption{Parameter values referring to the best-fit SYN++ models fitted to the spectra of SN~2021sjt. The coding is the same as in Table \ref{tab:local_20dpw}.}
\label{tab:local_21sjt} 
\begin{center}
\begin{tabular}{lccccc}
\hline
\hline
Ions & He I & O I & Si II & Ca II \\
\hline
  \multicolumn{4}{c}{SN~2021sjt 4d phase ($T_{\rm ph}~=~7$; $v_{\rm ph}~=~$15.5) } \\
\hline
$\log\tau$ & 0.4 & -0.4 & 0.0 & \\
$v_{\rm min}$ &15.5 & 15.5 & 15.5 &\\
$v_{\rm max}$ & 30.0 & 30.0 & 30.0 &\\
aux  & 2.0 & 1.0 & 3.0 & \\
$T_{\rm exc}$ & 5.0 & 8.0 & 16.0 & \\
\hline
 \multicolumn{4}{c}{SN~2021sjt 7d phase ($T_{\rm ph}~=~6$; $v_{\rm ph}~=~$15.0) } \\
\hline
$\log\tau$  & 0.5 & 1.5 & 0.2 & 2.0 \\
$v_{\rm min}$ &15.0 & 15.0 & 15.0  & 15.0\\
$v_{\rm max}$ & 30.0 & 30.0 & 30.0 & 30.0\\
aux &2.0 & 0.3 & 3.0 & 1.0\\
$T_{\rm exc}$  & 5.0 & 8.0 & 10.0 & 5.0 \\
\hline
 \multicolumn{4}{c}{SN~2021sjt 10d phase ($T_{\rm ph}~=~$5.8; $v_{\rm ph}~=~$13.0) } \\
\hline
$\log\tau$ & 0.2 & 1.0 & 0.3 & 2.6 \\
$v_{\rm min}$ &13.0 & 13.0 & 13.0 & 13.0 \\
$v_{\rm max}$ & 30.0 & 30.0 & 30.0 & 30.0\\
aux & 2.0 & 0.5 & 2.0 & 1.0 \\
$T_{\rm exc}$ & 5.0 & 5.8 & 10.0 & 5.0 \\
\hline
 \multicolumn{4}{c}{SN~2021sjt 11d phase ($T_{\rm ph}~=~$5; $v_{\rm ph}~=~$12) } \\
\hline
$\log\tau$  & 0.3 & 1.0 & 0.0 &\\
$v_{\rm min}$ &12.0 & 12.0 & 12.0 & \\
$v_{\rm max}$ & 30.0 & 30.0 & 30.0 & \\
aux  & 2.0 & 0.5 & 2.0  & \\
$T_{\rm exc}$ & 5.0 & 5.8 & 10.0 & \\
\hline
 \multicolumn{4}{c}{SN~2021sjt 19d phase ($T_{\rm ph}~=~$5; $v_{\rm ph}~=~$10) } \\
\hline
$\log\tau$  & 0.5 & 0.7 & 0.0 & 3.0 \\
$v_{\rm min}$ &10.0 & 10.0 & 10.0  & 10.0 \\
$v_{\rm max}$ & 30.0 & 30.0 & 30.0 & 30.0\\
aux  & 2.0 & 1.0 & 2.0 & 1.0 \\
$T_{\rm exc}$  & 5.0 & 15.0 & 10.0 & 10.0 \\
\hline
\end{tabular}
\end{center}
\end{table*}

\begin{table*}
\caption{The inferred physical parameters for SN~2020dpw and SN~2021sjt.}
\label{tab:tetaperve} 
\begin{center}
\begin{tabular}{lcccc}
\hline
\hline
Phase & T &  $v_{\rm ph}$ &  $\theta$ &  $\theta/v$ \\
(days) & (K) & (km s$^{-1}$) & ($10^8$ Mpc) & (Mpc day$^{-1}$) \\
\hline
\multicolumn{5}{c} {SN2020dpw} \\
\hline
16.92 & 9615 & 4454 &  2.492 &  0.647 \\
18.81 & 8401 & 4235 &  3.191 &  0.872 \\
19.80 & 8018 & 4128 &  3.195 &  0.896 \\
20.90 & 8532 & 4016 &  2.554 &  0.736 \\
30.85 & 7145 & 3192 &  3.210 &  1.164 \\
34.93 & 7017 & 2929 &  3.378 &  1.335 \\
35.83 & 6596 & 2876 &  3.523 &  1.418 \\
39.81 & 6722 & 2655 &  3.454 &  1.506 \\
40.80 & 6622 & 2604 &  3.399 &  1.511 \\
43.69 & 6628 & 2463 &  3.318 &  1.559 \\
44.69 & 6157 & 2417 &  3.772 &  1.807 \\
45.78 & 6674 & 2367 &  3.343 &  1.634 \\
49.66 & 5751 & 2204 &  3.798 &  1.995 \\
51.65 & 5721 & 2126 &  3.783 &  2.060 \\
53.74 & 5592 & 2048 &  3.881 &  2.193 \\
54.64 & 5782 & 2016 &  3.798 &  2.181 \\
55.63 & 5883 & 1981 &  3.845 &  2.246 \\
\hline
\multicolumn{5}{c} {SN2021sjt} \\
\hline
7.96 & 5134 & 14470 &  4.128 &  0.397 \\
10.25 & 4805 & 13296 &  4.181 &  0.436 \\
12.24 & 4673 & 12337 &  4.167 &  0.469 \\
14.13 & 4590 & 11466 &  4.241 &  0.514 \\
16.81 & 4210 & 10297 &  4.495  & 0.609 \\
18.11 & 3834 & 9764 &  4.516 &  0.639 \\
\hline
\end{tabular}
\end{center}
\end{table*}

\end{appendix}

\end{document}